\documentclass[a4paper,11pt]{article}
\usepackage{amsmath}
\usepackage{amssymb}
\usepackage{verbatim}
\usepackage{color}

\makeatletter

    \@addtoreset{equation}{section}
\makeatother
\newcommand{\vev}[1]{ \left\langle {#1} \right\rangle }
\newcommand{\cA}{{\cal A}}
\newcommand{\cB}{{\cal B}}

\newcommand{\cF}{{\cal F}}

\newcommand{\cN}{{\cal N}}

\newcommand{\cO}{{\cal O}}

\newcommand{\cZ}{{\cal Z}}

\newcommand{\ep}{\varepsilon}

\newcommand\be{\begin{equation}}
\newcommand\ee{\end{equation}}
\newcommand\nn{\nonumber}

\newcommand{\fpp}[1]{ \frac{\partial}{\partial {#1}} }

\def\l{\lambda}
\def\v{\varphi}

\def\L{\Lambda}

\newcommand{\ba}{\begin{eqnarray}}
\newcommand{\ea}{\end{eqnarray}}

\begin{document}


\begin{center}
{\LARGE \bf Irregular conformal block, spectral curve and flow equations}
\vskip 12mm
{\large  Sang Kwan Choi\footnote{email:hermit1231@sogang.ac.kr},
 Chaiho Rim\footnote{email:rimpine@sogang.ac.kr}
and Hong Zhang\footnote{email:kilar@sogang.ac.kr}
}\\
{\it Department of Physics and Center for Quantum Spacetime (CQUeST)}\\
{\it Sogang University, Seoul 121-742, Korea}
\end{center}

\vskip 12mm

\begin{abstract}

Irregular conformal block is motivated by the Argyres-Douglas type 
of N=2 super conformal gauge theory. 
We investigate the classical/NS limit of 
irregular conformal block 
using the spectral curve 
on a Riemann surface with irregular punctures, 
which is equivalent to the loop equation of irregular matrix model. 
The spectral curve is reduced to   
the second  order (Virasoro symmetry, $SU(2)$ 
for the gauge theory) 
and third order  ($W_3$  symmetry, $SU(3)$) 
differential equations 
of a polynomial with finite degree. 
The conformal and W symmetry 
generate the flow equations in the spectral curve 
and determine the irregular conformal block, 
hence the partition function of the Argyres-Douglas theory 
ala AGT conjecture.

\end{abstract}

\vskip 12mm

\setcounter{footnote}{0}

\section{Introduction}  

Irregular conformal block (ICB)  is closely related with Argyres-Douglas type (AD) 
of N=2 super conformal gauge theory in four dimensions \cite{AD}.  
AD type has the non-trivial infrared fixed point on the Coulomb branch
and does not allow marginal deformation.
Therefore, AD type of gauge theory is considered 
as a special class of super conformal gauge theory. 

According to AGT \cite{AGT}, 
the Nekrasov partition function \cite{Nekrasov1,Nekrasov2, NO} of the  gauge theory 
is equivalent to the conformal block of Liouville vertex operators
in  two dimensions. 
This connection is understood using the twisted compactification of the 
six dimensional ${\mathcal N}= (2,0)$ theory on a punctured Riemann surface
\cite{GMN_2009, BT_2009, NX_2010}. 
In this context, the Seiberg-Witten curve of the four-dimensional theory 
is identified with the spectral curve 
of Hitchin system on the Riemann surface with regular punctures.
The Hitchin system has simple poles and the residues are associated with the
mass parameter of the gauge theory \cite{H_1987, DW_1996}. 
On the other hand, the AD type theory is characterized in terms of irregular punctures,
poles of higher order \cite{CNV_2010, CV_2011}. Therefore, the irregular puncture 
is the key point to understand AD type theory. 

It is noted that  irregular punctures 
appear when the regular conformal block  
has the colliding limit \cite{{EM_2009}, GT_2012}. 
The colliding limit is a fusion of vertex operators 
so that multiple moments of 
Liouville charges of the vertex operators are present.  
It is like the collection of charges distributed over a small region
whose collection is viewed as an idealized system 
of total charge, dipole, quadrupole and multi-poles.
As a consequence, the  irregular punctures 
maintain the conformal symmetry 
but change the  conformal state.
Note that a regular puncture on the Riemann surface 
appears due to  a primary field,
which indicates the primary state, eigen-state of Virasoro generator $L_0$.
Similarly, the irregular puncture of order $(n+1)$ indicates an
irregular state of rank $n$,
which is a simultaneous eigen-state of positive Virasoro generators
$L_k$ with $n \le k \le 2n$. 
This irregular state is called  a Gaiotto state \cite{G_2009}
or a Whittaker state \cite{Whittaker}.  
Rank 0 state corresponds to the regular state.

According to AGT, the conformal block 
provides the partition function of gauge theory. 
In the same way, 
the partition function of AD type gauge theory is given by ICB. 
To find ICB, we will use the property of the  Penner-type random matrix model.  
The Penner-type matrix model is originally introduced 
to study the topological structure of the punctured Riemann surface \cite{P_1988}. 
It turns out that the Liouville conformal block is 
conveniently represented as the Penner-type matrix model, 
which is an equivalent way of writing 
of the Selberg integrals \cite{DV_2009, IO_2010}.

There are a few merits in using the matrix model. 
The Penner-type matrix model is easy to apply the colliding limit
which results in the irregular matrix model (IMM). 
From IMM one may obtain  ICB
by properly normalizing the partition function and compensating
$U(1)$ factor  \cite{CRZ_2015}. 
In addition, the random matrix model provides the loop equation, 
Ward-identity representing the conformal symmetry.
The loop equation allows one to investigate 
the detailed structure of the spectral curve corresponding to the Hitchin system. 
It turns out that the spectral curve contains 
flow equations corresponding to the conformal and W symmetry 
and fixes ICB and therefore, the partition function of AD type gauge theory
according to AGT.
It is remarkable 
that one may find ICB using only the conformal and W symmetry of the theory.  
This paper is mainly devoted to elaborating on the 
relation between ICB and spectral curve through flow equations. 
 
This paper is organized as follows. 
In section \ref{sec:virasoro}, we consider the case with $su(2)$ gauge group 
which has the Virasoro symmetry. 
We confine ourselves to the classical/NS (Nekrasov-Shatashivili) limit \cite{NS}
to present the main features of the system. 
NS limit is a limit where one of the 
deformation parameters $\epsilon_{1,2}$ of the guage theory 
goes to zero and the other (called $\epsilon$) remains to be finite. 
In the Liouville theory sense, 
the classical limit is achieved by 
taking the limit $\epsilon = \hbar Q$ finite 
where the scale parameter $\hbar$ goes to zero 
but the background charge $Q$ goes to infinite.  
One may resort to the classical/NS limit to get  simplified spectral curve
which still possesses the irregular punctures
and the Virasoro symmetry.  
This spectral curve reduces to the second order differential equation
of a polynomial of a finite degree $N$, size of the random matrix.  
In addition, we do not have the cut structure on the Riemann surface 
which usually appears at the large $N$ limit.
However, the concept of the filling fraction 
(identified with the Coulomb branch parameter in the gauge theory)
still remains and  plays an important role in ICB.  
We present the explicit form of the flow equations from the spectral curve 
and use the flow equations to find the the irregular partition function
and ICB. 
An explicit form of the irregular partition function and ICB 
are given for  lower rank cases.
 
In section \ref{sec:w3} we extend the idea 
to the case with $su(3)$ gauge group,
which has the Virasoro and $W_3$ symmetry.
IMM of $A_2$ is presented, 
which is obtained from the Toda field theory and its colliding limit. 
The spectral curve is obtained and flow equations are found 
by identifying the Virasoro and $W_3$ symmetry generator.  
Using a similar method, we construct partition function of IMM 
and ICB.  
In section \ref{sec:gauge}, we briefly summarize
the relation between gauge theories and the matrix model approach.
Section \ref{sec:summary} is the conclusion and outlook.
In the appendix, one can find the derivation of the loop equations of $A_2$ model  
(appendix \ref{app:loop}), 
representation of $W_3$ currents (appendix B),
and perturbative method to find the moments for the flow equation (appendix \ref{app:IMM}).

\section{Irregular conformal block with Virasoso symmetry}
\label{sec:virasoro}

\subsection{Irregular matrix model and spectral curve} 

The irregular matrix model with Virasoro symmetry is given 
as the $\beta$ deformed random matrix model 
\be
{Z}_{(m;n)}   =     \int \left( \prod_{I=1}^N d \lambda_I \right)
 \prod_{I < J} (\lambda_I - \lambda_J)^{2\beta} 
            e^{ \frac{\sqrt{\beta}}{g}  \sum_I V(\lambda_I)}\,,
\label{eq:partition-mn}
 \ee
whose  potential has the form 
\be 
 V  (z )
=   c_0 \log z - \sum_{k=1 }^{n}   \left( \frac {{  c_k} } {k z^{k}} \right)   
+ \sum_{\ell=1}^{m}   \left( \frac { c_{-\ell}~ z^{\ell}}   {\ell}  \right )  \,.
\ee
The  deformed parameter $\beta$ is  related 
with the Liouville screening charge $b= i \sqrt{\beta}$.
In addition,  a small expansion parameter $g$ is introduced in the partition function,
which is equivalent to the Liouville scaling parameter $ \hbar =- 2 i g $; 
 $\sqrt{\beta} /g = - 2 b /\hbar$.  
It is noted that the parameter $c_k$ in the potential is 
the one denoted as $\hat c_k$ 
in \cite{NR_2012}. We simplify the notation by deleting hat 
and therefore, $c_k$ stands for old  $\hbar c_k$. 

The partition function \eqref{eq:partition-mn} 
is obtained from the colliding limit 
of regular $(m+n+2)$-point conformal block.
As a result, the potential parameter $c_k$ 
is related with the   Liouville charge $\alpha_a$
and the  vertex operator position $z_a$
as  $   c_k= \sum_{r=1}^{n} \hbar \alpha_r(z_r)^k  $ with $k> 0$
and  $   c_{-\ell}=- \sum_{a=1}^{m} \hbar \alpha_a(z_a)^{-\ell}$ with $\ell>0$.
One may regard the partition function as the  irregular correlation
of two irregular vertex operators of rank $m$ and $n$,
considering the rank-$n$ irregular operator as 
$I_c^{(n)}(z) = e^{2 \Phi(z)}$
with $\Phi(z) = \sum_{k=0}^n  c_k   \partial_k \varphi /k! $
where $ \partial_k \varphi $ stands for  $k$-th derivative of the Liouville field. 
In this sense,  the rank-0  irregular operator
reduces to the regular vertex operator  
$V_\alpha (z) =e^{2 \alpha \varphi(z)} $.
In addition, the two -point irregular correlation 
is viewed as the inner product of 
  irregular states of rank $m$ and $n$.
Indeed, the two-point correlation
is identified with the inner product
if one  takes care of the  proper normalization of the state  
and $U(1)$ factor which arises at the colliding limit \cite{CRZ_2015}. 
Multi-point irregular conformal block is also 
constructed in a straight-forward way. 

The partition function has the conformal symmetry 
and the symmetric property is encoded in the loop equation,
\be
4 W(z)^2 + 4 V'(z) W(z) + 2  \epsilon  W'(z)    =f(z) . 
\label{eq:loop}
\ee
Here $ W(z)$ is the resolvent,   
$ W(z) = g \sqrt{\beta}   \left\langle\sum_I  \frac{1} {z - \lambda_I} \right\rangle$
where the bracket $  \left\langle \cdots  \right\rangle$ 
denotes the  expectation value with respect to the matrix model.  
$  f(z)    =   4g \sqrt{\beta} \left\langle
 \sum_I \frac{V'(z)- V'(\lambda_I)}{z - \lambda_I} 
\right\rangle $.
The loop equation \eqref{eq:loop} is given at the classical/NS limit.
Note that the classical/NS limit is obtained when $\hbar \to 0$ 
and $b \to \infty$ so that $\epsilon = \hbar b $ is finite
and  multi-point resolvent contribution vanishes \cite{MMM}. 
Therefore, we ignore the two-point resolvent in the original loop equation. 

From the Liouville field theory point of view,
the large  $b$ limit is the same as the one 
with the small $b$ limit because of the  duality $b \to 1/b$
is present in  of the Liouville theory.\footnote{Note that before taking the classical/NS limit 
 the parameter $\epsilon $ in the loop equation 
 is $ \hbar Q = 2 g(\sqrt{\beta} - 1/\sqrt{\beta})  $
where $Q=b+1/b$ is the background charge of the Liouville theory.
Thus, the original $\epsilon $  is  invariant under the $b \to 1/b$ duality. }
Therefore, the same finite  $\epsilon= \hbar Q$ is obtained 
either at the NS limit  or at the classical limit ($b\to 0$) of the Liouville theory.   
It is also noted that the same duality also appears in the corresponding gauge theory.
The $\Omega$ deformation parameters  $\epsilon_1$ and $\epsilon_2 $
are identified as  $\epsilon_1= \hbar b$ and $\epsilon_2= \hbar /b$
according to to AGT conjecture 
and  $\epsilon = \epsilon_1+ \epsilon_2$. 
The $b \to 1/b$ duality corresponds to  $\epsilon_1 \to \epsilon_2$ duality. 
If one of the deformation parameter vanishes while the other remains finite, 
the limit is called Nekrasov-Shatashivili (NS) limit,
which is the same as the classical limit of the Liouville theory
or the matrix model. 
 
The loop equation \eqref{eq:loop} can be put  in a more informative form
if one uses  $ x =  2 W  +  V'  $,
\be
x ^2  + \epsilon x'    +  \xi_2(z) =0 \,,
\label{virasoro-spectral-curve} 
\ee 
which can be regarded as a spectral curve  
with $z$ the (complex) spectral parameter. 
The analytic structure of the spectral curve  is specified  by  
$  \xi_2 (z) = - V'^2 +\epsilon  V'' - f $
which  has a pole of order $2n$ and a zero of order $2m$
on the Riemann surface:   
\be
 \xi_2(z) =\sum_{k=-2m}^{2n} \frac{\Lambda_k}{z^{k+2}}
-\sum_{a=-{m}}^{n-1} \frac{d_a}{z^{a+2}}  \,,
\label{eq:xi2}
\ee 
where $ f = \sum_{a=-{m}}^{n-1} \frac{d_a}{z^{a+2}} $ is used. 
The spectral curve indicates that  one may view the mode of $\xi_2(z)$ as the 
conserved quantity appearing in the integrable theories \cite{H_1987, DW_1996}. 
Indeed, if one ignores $f(z)$ (putting $d_a=0$), 
$\xi_2(z)$ is given in terms of $ \Lambda_k$  
which is a constant:
\be
\Lambda_k= \epsilon(k+1)  c_k 
-\sum_{r+s=k} c_r   c_s  \,.
\ee 

The new feature in the loop equation \eqref{virasoro-spectral-curve} is
that $f(z)$ has the special role in finding the partition function \cite{NR_2012}. 
This role is closely related with the Virasoro symmetry. 
Note that $\xi_2(z) $ is identified with the expectation 
value of the energy momentum tensor  (Virasoro current)
\be
  \xi_2 (z)=  \langle T(z) \rangle 
=\sum_{k \in \mathbb{Z}} \frac{  \left\langle  L_k \right\rangle }{z^{k+2}} \,.
\ee 
Comparing the two, one notes that 
$\Lambda_k$ is the eigenvalue of the Virasoro mode 
$L_k$ for $n \le k \le 2n$ and  $-2m \le k \le -m$.
The positive mode  $L_k$ with  $n \le k \le 2n$ applies on the ket 
so that  the  ket  is the simultaneous eigenstate 
of $L_k$ ($n \le k \le 2n$). 
The negative mode applies on the bra 
since $L_k^\dag = L_{-k}$
and therefore, the bra is the simultaneous eigenstate 
of $L_k$   ($-2m \le k \le -m$).  

However, the extra mode  $\Lambda_a + d_a$ ( $-m  \le a \le n-1$)
present in   $\xi_2(z)$  is not a simple constant due to $d_a$
and therefore,  not an eigenvalue. 
Instead, the mode represents the  expectation value  of $L_a$.
It is noted that  $d_a$  is directly related to
the partition function  or free energy 
$F_{(m;n)}=-\hbar^2\log {Z}_{(m;n)}$ \cite{CR_2013}: 
\be
d_a= 
\begin{cases} 
& v_a (F_{(m;n)} )    
\quad ~~~~~~~~~~~~~~
{\rm if}~0\le k \le n-1 
\\   
&v_a  (F_{(m;n)}  )  + 2 \epsilon  N  c_k
\quad ~\,~{\rm if}~-m < a <0
\\ 
&  2 \epsilon  N c_{-m}
\quad ~~~~~~~~~~~~~~~~~
{\rm if}~a=-m \,,
\end{cases}
\label{eq:da-W2}
\ee  
where $v_a$ is given as derivative with respect to the parameters $c_k$'s:
\be
v_a=
\begin{cases}
& \sum_{s>0}  \,s  \left( c_{s+a} \fpp{ c_s} \right)  
\quad ~~~~ {\rm if}~0\le k \le n-1
\\ 
&
\sum_{s<0}\, (-s) 
 \left(  c_{s+a} \fpp{ c_s} \right)  
\quad {\rm if}~-m < a <0\,.
\end{cases}
\label{eq:va-W2}
\ee   
In fact, $v_a$ in  \eqref{eq:va-W2} 
is  the  representation of  Virasoro mode  $L_a$ 
on the parameter space
and $d_a$ is  the vector flow $v_a$ of the partition function. 
It is, however, noted that 
$v_a$ behaves differently,
depending on the sign of  $a$. 
When  $a \ge 0$,
$ v_a$ acts on the parameter space $\{c_1, \cdots, c_n\}$ of the right state  
and satisfies the commutation relation
$ [ v_k , v_\ell] = -(k-\ell) v_{k +\ell} $.
On the other hand, 
$ v_a$  with $a < 0$ acts on the parameter space
$\{ c_{-m}, \cdots, c_{-1} \}$  of the left state 
and satisfies the commutation relation
$ [ v_k , v_\ell] = (k-\ell) v_{k +\ell} $. 

The appearance of left/right representation of $L_k$ 
brings an unpleasant feature. 
Suppose we evaluate  ${Z}_{(0;1)}$
whose parameter space is $\{c_1\}$.
The vector flow $d_0$ of $v_0$ can determine the partition function 
using \eqref{eq:da-W2}.  
In fact, the partition function is easily obtained 
if one scales $\lambda_I \to c_1 \lambda_I$ 
in \eqref{eq:partition-mn}.
This shows that the flow equation \eqref{eq:da-W2}
contains the information of the parameter scaling. 
On the other hand,  suppose one evaluates ${Z}_{(1:0)} $ 
whose parameter space is $\{c_{-1} \}$.
The partition function is  found if one scales 
$\lambda_I \to  \lambda_I/c_{-1} $.  
However, there is no $v_{-1}$ representation 
in \eqref{eq:da-W2} when $m=1$
and we cannot fix the partition function using the flow equation. 
This unpleasant feature does not raise any  problem 
in evaluating ${Z}_{(m;n)} $ ($m \ne 0$)
if one scales  $\lambda_I^m \to   \lambda_I^m/c_{-m} $ first.  
(See more details in subsection \ref{sec:partition-0n}). 

Nevertheless, this unpleasant feature can be avoided if  
one takes into account of the conformal 
invariance of the partition function.  
Note that conformal transformation $\lambda_I \to 1/ \lambda_I$
changes the parameter space $\{ c_{-m}, \cdots, c_{-1}, c_1, \cdots, c_n; c_0\}$  
into its dual space \\
$\{ \bar c_{-n}, \cdots, \bar  c_{-1}, \bar  c_1, \cdots, \bar c_m; \bar  c_0 \}$
where $\bar c_k = -c_{-k}$ ($k \ne0$)
and $\bar c_0 = c_\infty$.
Here $c_\infty$ is determined by  the neutrality condition 
\be
c_0 + c_\infty + N \epsilon  = \epsilon\,.
\label{eq:neutrality-W2}
\ee
The neutrality condition is hidden in the matrix model 
but is manifest in the Liouville conformal block.
The conformal block is  evaluation using the perturbation 
with the insertion of $N$ screening operators
and the neutrality condition is required 
$ \sum_a \alpha_a  + N b = Q$ 
so that the conformal block is non-vanishing.   
At the colliding limit, the neutrality condition reduces to  
$ c_0 + c_\infty + N \hbar b = \hbar Q $
where $c_0= \sum_{r=1}^{n} \hbar \alpha_r $ 
is the total Liouville charge  (scaled by $\hbar$) at the origin 
and $c_\infty $ is the charge at infinity. 
If one has the classical/NS limit,  
one has the neutrality condition \eqref{eq:neutrality-W2}. 

Using this conformal symmetry, we have two left representations
instead of one left and one right representations:
One is the original representation 
$v_a$ with $a \ge 0$ which applies to original parameter space
$\{c_1, \cdots, c_n \}$  
with parameters $\{ c_{-m}, \cdots, c_{-1}; c_0 \}$  intact. 
The other is $\bar v_a$ with $a \ge 0$ 
which applies to the dual parameter space
$\{ \bar  c_{-n}, \cdots, \bar c_{-1}\}$
with 
$\{ \bar  c_1, \cdots, \bar c_m; \bar  c_0 \}$
intact. 
Here the barred  representation  $\bar  v_a$ is defined in
the same form as in \eqref{eq:va-W2} 
but with $\{c_k \}$ replaced with $\{\bar  c_k \}$
and commutes with $v_a$; $[v_a, \bar  v_b]=0$.
The barred partition function is the same as the original one 
because of the conformal invariance of the partition function.  
Therefore, one may find the partition function from
the flow equation presented in the spectral curve
and the existence of the partition function is guaranteed 
due to the consistency condition of $d_a$ and its dual $\bar  d_a$: 
\begin{align}
& v_a (d_b) - v_b (d_a) = -(a-b ) d_{a +b }  
\nn \\ 
& \bar v_a ( \bar d_b) - \bar v_b ( \bar d_a) = -(a-b ) \bar d_{a +b } 
\nn
\\
&   v_a ( \bar d_b) - \bar v_b (   d_a) = 0
 \,.
\label{eq:consistency}
\end{align}   
It is simple to note that
if one replaces $d_a$ with  $\Lambda_k$,
then $\Lambda_k$ also satisfies the consistency condition
\eqref{eq:consistency}. 
 
If one uses the flow equation  to construct the partition function,
the major step is to find the values of  $d_a$ and $\bar d_a$
directly from the  analytic property of the spectral  
curve \eqref{virasoro-spectral-curve}.
In the usual large $N$ approach \cite{C_2010, CEM_2011},
one expands the spectral curve in powers of  $\hbar$  
keeping $\hbar N= O(1)$ and assuming $x $ is $ O(1)$.
In this case, $\hbar Q$ is the sub-dominant order of $\hbar$ \cite{NR_2012}
and the spectral curve is given as  $ x  = \pm  \sqrt{- \xi_2} $ at the leading order.
The solution results in $(m+n)$  square-root branch cuts
and provides the double covering 
of the Riemann surface \cite{DV_2009, DV_2002}.
The contour integral of $x$ over a certain cut becomes 
an elliptic integral 
and is identified with the filling fraction. 
In this way, one can find  $d_k$ in terms of the filling fraction 
and the parameters $c_k$'s. 
Once $d_k$ is known, the free energy is constructed according to 
\eqref{eq:da-W2} and \eqref{eq:va-W2}.
This procedure is very interesting from the gauge theory point of view. 
Since the filling fraction is identified with the Coulomb branch parameter 
of the gauge theory according to AGT 
and the spectral curve is the Seiberg-Witten curve of the Hitchin system,
Seiberg-Witten curve determines the partition function 
of the  Argyres-Douglas  gauge theory  with the parameters $\{c_k\}$
which can be rewritten in terms of masses and Coulomb branch expectation value 
of the gauge theory. 

At the classical/NS limit, the procedure does not change much. 
Still, there appear important modifications. 
There are no branch cuts in the spectral curve: 
Only  $N$-number of simple poles are present. 
This fact can be seen as follows. 
Let us consider  an  expectation value  
\be
P(z) \equiv   \left\langle  \prod_I  ( z-\lambda_I ) \right\rangle   
= \sum_{A=0}^N  P_A  z^A  = \prod_\alpha (z-z_\alpha)
\label{eq:PA}
\ee 
which is a polynomial of degree $N$ with  $P_N=1$.
$z_\alpha$'s are $N$-zeros  of the polynomial.
We assume all zeros are distinct.
Note that $P(z)$ is related with the resolvent $W(z)$ at the classical/NS limit 
since
\be 
\log \left( \frac { P (z)}{ P (z_0)} \right) 
 =  \frac2 {\epsilon }  
\int_{z_0}^z      dz' ~W(z' )  \,.
\label{eq:P-W}
\ee 
Here we use the fact the multi-point resolvent contribution vanishes at the classical/NS limit
\cite{MMM, BMT_2011}.
Taking the derivative of \eqref{eq:P-W},
 one has  $ 2 W(z)  = \epsilon  { P' (z)}/ { P (z)}$
which can be put as 
\be
2 W(z)    =  \epsilon \sum_{\alpha=1}^N \frac 1 {z-z_\alpha}\,.
\ee 
Therefore, only  $N$-simple poles appear in the spectral curve,
which substitute the cuts present at the large $N$ expansion.  

The monic polynomial $P(z)$ has the central role in finding $d_a$. 
Explicitly, the spectral curve \eqref{virasoro-spectral-curve}  
reduces to the second order differential equation of  $P(z) $:
From \eqref{eq:loop} one has 
\be
{\epsilon}^2  P''(z)   +  2{\epsilon} V'(z) P'(z)  = f(z) P(z)\,.
\label{eq:P}
\ee 
One may expand the differential equation 
in power series of $z$ and find 
finite number of algebraic relations of $P_A$'s and $d_a$'s
since $P(z)$ is the polynomial of degree $N$. 
Therefore, one can find $d_a$'s using the algebraic relations only.  
It should be noted that the solution is not unique.
The solution depends on how the zeros of the polynomial $P(z)$ 
distribute around the stationary point of the potential.
Therefore, the filling fraction may be applied   
to the distribution of the zeros 
even though there are no branch cuts on the Riemann surface.    
With the filling fraction one may fix the relevant solution of \eqref{eq:P}.
We provide some explicit examples in the next two subsections. 

\subsection{Partition function ${Z}_{(0;n)}$ } 
\label{sec:partition-0n}

In this section we first consider the case where the potential 
is given as logarithmic and inverse polynomials only: $c_k =0$ when $k <0$.
The partition function with this potential 
is regarded as two point correlation between one regular 
and one irregular vertex operator. 
In this case, $d_0$ is simply obtained 
if one uses large $z$ expansion of the loop equation:
\be
d_0= 2 \epsilon c_0 N + \epsilon^2 N(N-1) \,.
\ee
However, other values of $d_k$ are not easy to find. 

Suppose one evaluates the simplest partition function, rank $n=1$. 
In this case,  $d_0$ is enough to find 
the partition function if one uses the flow equation \eqref{eq:da-W2}.  
${ Z}_{(0;1)} = c_1^{-d_0/\hbar^2} \cN_{(0;1)} $ 
where $\cN_{(0;1)} $ is the normalization independent of $c_1$.
The same result is obtained if one rescales $\lambda_I$ 
in the partition function \eqref{eq:partition-mn}.
In addition, one can find $P(z)$ explicitly since 
the mode expansion gives a recursive relation:
$ P_A= \xi_{A+1}  P_{A+1} $  
where $\xi_{A+1}=2 (A+1)\epsilon c_1/  (d_0 -2 \epsilon c_0 A - \epsilon^2 A(A-1) )$.
Here  $P_{N+1} =0$ and $\xi_{N+1}=0$ are used 
for notational simplicity and  $P_N=1$.
The recursion relation shows that  for $0 \le A <N$ 
\be
P_A=\prod_{\alpha=A+1}^N   \xi_{\alpha}\,.
\ee 
Suppose $N=1$, one has $P(z) =z-z_1$ and 
$P_0 =\xi_1 =c_1/c_0= -z_1$. 
The zero $z_1$ corresponds to the stationary point of the potential $V$.
When $N>1$, the solution  $P_A$ provides 
the information of  $N$ zeros around the stationary point
since  $P_A$ is  written as the the polynomial of  $N$-zeros 
in a permutation invariant form:
$P_A= \sum_{\{\alpha_i\}} z_{\alpha_1} \cdots z_{\alpha_{N-A}}$ 
where the index sum is ordered.

If the rank is  greater than one,  
the solutions $P_A$ and $d_k$ 
are more complicated. 
To get the idea on this solution, let us consider the rank two with  $N=1$.
In this case we need $d_0$ and $d_1$. 
$d_0$ is trivially given: $d_0 = 2 \epsilon c_0$. 
On the other hand, one has $d_1 = 2 \epsilon c_1 -d_0 P_0$ 
and $d_1$ is fixed by a quadratic equation:
\be
d_1^2  -2  \epsilon c_1 d_1 + 2 \epsilon c_2 d_0  =0\,.
\ee
One has two solutions: 
$d_1^\pm = \epsilon c_1 \left( 1 \pm \sqrt{1-\eta} \right) $
where $\eta= 4c_2c_0/c_1^2$.
Note that two solutions shows that  
$P_0^-  \sim  c_1/c_0$  and $P_0^+ \sim c_2/c_1$,
each of which lies near one of two stationary points of the potential. 
This two different solutions are due to the fact
that zeros of the polynomial 
(or the poles of the resolvent) 
distribute around 
two different stationary points of the potential.
One may classify the solutions in terms of the filling fraction. 
Therefore, if one integrates the flow equation \eqref{eq:da-W2},
one  has the different free energy depending on the filling fraction,  
\be
F_{(0;2)}^\mp= \epsilon c_0 \left(  \log c_2 
-\frac18 \int^\eta   \frac {1 \mp \sqrt{1 - x }}{x^2}  dx \right)\,.
\ee

For general $N$, one can easily convince that
$P_{N-1} $ has $N+1$ solutions.
The solutions  correspond to the zero distribution 
of $P(z)$ so that $N= N_1 + N_2 $   
where $N_1$ zeros at one stationary point and $N_2$ at the other.  
However, it is not easy to find the exact form of $d_a$ when $N$ is large
and one may resort to perturbative expansion. 
One way to find the solution is using the power series of $\epsilon$
with $\epsilon$ small.
Note that $W(z) = O(\epsilon)$ and $f(z) =O(1)$ 
while $V(z) = O(1)$ from the definition. 
In this case, we may apply  the $\epsilon$ expansion 
to the loop equation \eqref{eq:loop} directly.
Denoting $W(z)=\sum_{k \geq 1} \epsilon^k W^{(k)}(z)$
and $f(z)=\sum_{k \geq 1} \epsilon^k f^{(k)}(z)$,
one has the leading order contribution 
\be
 2W^{(1)}(z)=\frac{f^{(1)}  }{2 V' } \,.
\ee
This shows that the poles of the resolvent 
is located closely to the stationary points of the potential
at the leading order.    
This is consistent with the expectation that 
zeros of the polynomial $P(z)$ are accumulated
around the stationary points. 
Denoting $N_k$  for the filling fraction 
around  the stationary point $\xi_k$ of the potential, 
one obtains the identity 
\be
N_k=\oint_{\cA_k} \frac{f^{(1)}}{2 V'} dz
\label{epsilon^1}
\ee
where $\cA_k$ is the contour encircling $\xi_k$ only. 
Since the number of stationary point is the same 
as the number of $d_a$, one can fix  $d_a$ 
at $ O(\epsilon^1)$.
At the second order of the loop equation, 
one has the identity 
\be
0=\oint_{\cA_k} dz
\left\{
\frac{f^{(2)}}{2 V'}-\frac{(f^{(1)})'}{4 (V')^2}
+\frac{(2 V''-f^{(1)}) f^{(1)}}{8 (V')^3} 
\right\}   \,.
\label{epsilon^2}
\ee
Here the identity assumes that the contour  $\cA_k$
encircles all the poles of the resolvent corresponding to the fractional 
number $N_k$. In this way, one finds $d_a$ order by order in $\epsilon$
and therefore, the partition function. 

For the case rank 2 we have  $d_1$:
\be 
\begin{split}
d_1&=-2 \epsilon c_0( N_2 \xi_1+N_1 \xi_2) 
\\
&+\epsilon^2 \frac{\xi_1 \xi_2}{(\xi_1-\xi_2)^2}
\left[ N_2(N_2-1)(2 \xi_1-\xi_2)-N_1(N_1-1)(\xi_1-2 \xi_2)
-2 N_1 N_2(\xi_1+\xi_2) \right]
+\cO(\epsilon^3)
\end{split}
\ee
where $\xi_1$, $\xi_2$ are two stationary points
satisfying $c_0 z^2+ c_1 z+c_2=c_0(z-\xi_1)(z-\xi_2)$.
Using the flow equations $d_0=v_0 \left( F_{(0:2)} \right) $ 
and $d_1=v_1 \left( F_{(0:2)} \right) $, 
we get the free energy $F_{(0:2)}$. 
The partition function is given up to $\cO(\epsilon^2)$,
\be
\begin{split}
{Z}_{(0;2)}&= c_2^{-\frac{\epsilon}{\hbar^2} c_0 (N_1+N_2)
-\frac{\epsilon^2}{\hbar^2} \frac34 (N_1(N_1-1)+N_2(N_2-1))}
 \left(c_1^2-4c_0 c_2 \right)
^{\frac{\epsilon^2}{4 \hbar^2} (N_1(N_1-1)+N_2(N_2-1)-4 N_1 N_2)}
\\
&\times
\left(\frac{c_1+\sqrt{c_1^2-4c_0 c_2}}
{c_1-\sqrt{c_1^2-4c_0 c_2}}\right)^{-\frac{\epsilon}{\hbar^2} c_0 (N_1-N_2)
-\frac34 \frac{\epsilon^2}{\hbar^2} (N_1(N_1-1)+N_2 (N_2-1))}
\\
&\times
e^{-\frac{\epsilon}{\hbar^2} \frac{c1}{2c_2} \left( \sqrt{c_1^2-4c_0 c_2} (N_2-N_1)
+c_1 (N_1+N_2) \right)}
\end{split}
\ee

For the rank $n$ with $N$ zeros, one has 
$ \frac{(N+n-1)!}{N!(n-1)!}$ solutions. 
One may view this solutions as the zero distribution 
with $N=\sum_{i=1}^n N_i$ with $N_i$ zeros around 
each stationary point of the potential.   
Considering this zero distribution, 
one may expect that the solutions can be obtained perturbatively 
using parameter ratio set 
$\{\frac{c_1}{c_0}, \frac{c_2}{c_1}, \cdots, \frac{c_n}{c_{n-1}} \}$.
Each ratio stands for each stationary point.  
The same  conclusion also holds for 
${ Z}_{(n;0)} = \bar  {Z}_{(0;n)}$
if one works with the barred notations.

\subsection{Partition function ${ Z}_{(m;n)}$ and ICB} 

If one considers the case rank $(m;n)$, 
the potential contains finite Laurent series. 
One may evaluate the flow equations by finding 
$d_a$'s. 
The special feature of the  partition function ${Z}_{(m;n)}$ 
is that one may evaluate ICB 
of irregular vertex of rank $m$ and $n$. 
In this case, ICB is identified 
with the inner product $\langle I^{(m)}| I^{(n)} \rangle$ 
of two irregular states  
of rank $m$ and $n$. 
The relation of ICB with  ${Z}_{(m;n)}$
is given in  \cite{CRZ_2015}.
\be
{\cal F}_\Delta^{(m:n)}  (\{c_{-k }\} :\{ c_k \})= 
\frac{e^{\zeta_{(m:n)}}   
{Z}_{(m;n)}  (c_0 ; \{c_{\ell}  \} )} 
{Z_{(0:n)}  (c_0 ; \{c_k \} )  Z_{(0:m)}   (\bar c_0 ; \{\bar c_k \} )) }\,. 
\label{Virasoro_ICB}
\ee   
 $Z_{ (m;n) } $ is divided by  $ Z_{(0;n)} $ and $Z_{(0;m)} $ 
to give the proper normalization.
The extra factor $e^{\zeta_{(m:n)}}$ 
comes when $m$ vertex operators 
are put at the infinity and  $n$ operators at the origin.  
Explicit result is given as  $\hbar^2 \zeta_{(m;n)} 
= -\sum_k^{{\rm min}(m,n)} 2 (c_k c_{-k})/ k$. 
Therefore, it is obvious that ICB is exponentiated 
and the exponent should be inversely   
proportional to $\hbar^2$  \cite{RZ_2015},
considering the $\hbar$ dependence  
of the free energies and  $\zeta_{(m;n)}$.

For example, ICB ${\cal F}_\Delta^{(1;1)}$  
is obtained directly using the relation \eqref{Virasoro_ICB}.
$Z_{(0;1)}(c_0;c_1)$ can be easily obtained from the flow equation
$v_0 \left( F_{(0;1)} \right)=d_0$ and 
its dual $Z_{(0;1)}(\bar c_0;\bar c_1)$: 
\be
Z_{(0;1)}Z_{(0;1)}= c_{-1}^{(\epsilon N(\epsilon N+2c_0)
-\epsilon^2 N)/\hbar^2} \,
\eta_0^{-(\epsilon N_0(\epsilon N_0+2c_0)
-\epsilon^2 N_0)/\hbar^2}  
\ee
where $\eta_0 \equiv c_1 c_{-1}$.
To evaluate  $Z_{(1:1)}$, we first obtain 
$c_{-1}$ dependence by  rescaling  
$\lambda_I \to \lambda_I/c_{-1}$: 
$Z_{(1;1)}= c_{-1}^{(\epsilon N(\epsilon N+2c_0)
-\epsilon^2 N)/\hbar^2} \widetilde{Z}_{(1;1)}$.
$\widetilde{Z}_{(1:1)}$ is the partition function
with the potential $\widetilde{V}(z)=c_0 \log z+z-\eta_0/z$,
which is to be evaluated from the flow equation using 
$d_0=v_0 \left(-\hbar^2 \log \widetilde{Z}_{(1;1)} \right)$.
We use the $\epsilon$ expansion 
to obtain the parameter dependence of $d_0$:   
\be
d_0=-2 \epsilon (N_0 \xi_1+ N_\infty \xi_2)
+\epsilon^2 \frac{
N_0 (N_0-1) \xi_1^2-4 N_0 N_\infty \xi_1 \xi_2
+N_{\infty}(N_\infty-1) \xi_2^2}{(\xi_1-\xi_2)^2}+\cO(\epsilon^3)
\ee
where $\xi_1$ and $\xi_2$ are stationary points of $\widetilde{V}(z)$. 
The flow equation shows that 
\be
\begin{split}
Z_{(1;1)}&= c_{-1}^{\frac1{\hbar^2}(\epsilon N(\epsilon N+2c_0)
-\epsilon^2 N)} \, \eta_0^{-\frac{\epsilon^2}{2\hbar^2}
 (N_0^2-N_0+N_\infty^2- N_\infty)}
\left(\frac{c_0+\sqrt{c_0^2-4\eta_0}}{c_0-\sqrt{c_0^2-4\eta_0}}\right)
^{-\frac{\epsilon^2}{2\hbar^2} (N_\infty^2- N_\infty-N_0^2+N_0)}
\\
& \times
\left(c_0-\sqrt{c_0^2-4\eta_0}\right)^{-(2\epsilon c_0 N_0)/\hbar^2}
\left(c_0+\sqrt{c_0^2-4\eta_0}\right)^{-(2\epsilon c_0 N_\infty)/\hbar^2}
\\
& \times
\left(c_0^2-4\eta_0\right)
^{-\frac{\epsilon^2}{4\hbar^2} (N_0^2-N_0-N_\infty^2+ N_\infty+4 N_0 N_\infty)}
e^{-\frac{2 \epsilon}{\hbar^2} (N_0-N_\infty) \sqrt{c_0^2-4\eta_0}}\,,
\end{split}
\ee
up to $\cO(\epsilon^2)$.
From the result, one can obtain $\cF_{\Delta}^{(1;1)}$.
We provide $\cF_{\Delta}^{(1:1)}$ in powers of $\eta_0$ and $\epsilon$:   
\be
\begin{split}
\cF_{\Delta}^{(1;1)}&=\left[1-2\frac{\eta_0}{\hbar^2}+2\frac{\eta_0^2}{\hbar^4}
+\cO(\eta_0^3) \right]
+\epsilon \left[ \frac{2(N_0-N_\infty)}{c_0} \frac{\eta_0}{\hbar^2}
-\frac{(4 c_0^2- \hbar^2)(N_0-N_\infty)}{c_0^3} \frac{\eta_0^2}{\hbar^4}
+\cO(\eta_0^3) \right]
\\
&-\epsilon^2 \bigg[
\frac{2(N_0^2-N_0-2 N_0 N_\infty)}{c_0^2}\frac{\eta_0}{\hbar^2}
+\bigg\{\frac{\hbar^2 \left(7 N_0(N_0-1)
+N_\infty(N_\infty-1)-16 N_0 N_\infty \right)}{2c_0^4}
\\
&\qquad ~~~ -\frac{2\left(N_0(3N_0-2)+N_\infty^2-6 N_0 N_\infty\right)}
{c_0^2} \bigg\}\frac{\eta_0^2}{\hbar^4}+\cO(\eta_0^3) \bigg]
+\cO(\epsilon^3)
\end{split}
\label{(1:1)ICB}
\ee

One may also evaluate ICB directly using  the perturbative approach 
of IMM as noted in \cite{CRZ_2015}.  
This is because the relation \eqref{Virasoro_ICB}
shows that one needs  $Z_{ (m;n) } $ 
with proper normalization and compensates by the $U(1)$ factor.  
For the perturbative approach one may divide 
the potential matrix with $N$ eigenvalues into 
the one with $N_0$ eigenvalues
and the other with  $N_\infty$
so that $N=N_0+N_\infty$.
The normalization  $Z_{ (0;n) } $  is obtained 
if one uses $ V_{(0:n)} $ 
instead of  $ V_{(m:n)} $   with $N_0$ eigenvalues in  $Z_{ (m;n) } $.
Therefore, one may consider the potential $V_{ (m;n) } $ 
with $N_0$ eigenvalues composed of $ V_{(0:n)} $ and the rest 
so that $ V_{(0:n)} $ is treated as the reference potential $V_0 $
and the rest  as the perturbative one $\Delta V _0$: 
\be 
V_0 
= \sum_{I=1}^{N_0} \Big( c_0 \log \lambda_I - 
 \sum_{k=1}^{n} \frac{c_k} {k  \lambda_I^{k}} \Big)
\,,~~~~~
 \Delta V _0  =
  \sum_{I=1}^{N_0} \Big(
  \sum_{\ell=1}^{n} \frac { c_{-\ell}}{\ell}  { \lambda_I^{\ell}} \Big) \,.
\ee

Likewise, the potential $V_{ (m;n) } $ with $N_\infty$ eigenvalues
composed of $V_{ (m;0) } $  as the reference potential  $V_\infty$  and the perturbation  $\Delta V_\infty$
The partition function may be rewritten using the conformal 
transformation  $\lambda_J \to 1/ \mu_J$
so that one can use the dual form   
\be 
 V_\infty =  \sum_{J=1}^{N_\infty} \Big(
 \bar{c}_0 \log \mu_J  - \sum_{\ell=1}^{m} \frac { \bar c_{\ell}} {\ell\mu_{J} ^{\ell}}  
 \Big)
\,, ~~~~~
\Delta V _\infty = \sum_{J=1}^{N_\infty} \Big(
 \sum_{k=1}^{n} \frac {\bar c_{-k}} {k}  \mu_{J}^k  \Big) \,.
\ee

Once the normalization is done, the cross term $\prod_{I, J}  (1 - \lambda_I \mu_J)^{2\beta}  $ remains from the Vandermonde determinant. 
Therefore, the perturbative approach is to evaluate the expectation value 
$ \left\langle \prod_{I, J}  (1 - \lambda_I \mu_J)^{2\beta}  
 \exp \Big( 
\frac{\sqrt{\beta}}g (\Delta V_0 (\lambda_I) + \Delta V_\infty (\mu_J) ) 
\Big)   
\right\rangle_{\!\! 0}
$
where $ \left\langle \cdots \right\rangle_{\! 0}$
refers to the expectation value with respect to the reference potential. 
Practically, one may evaluate the expectation values
from the loop equations of  the partition functions $Z_{(0:n)}$ and $Z_{(0:m)}$
by using the large $z$ expansion of the resolvents. 
However, this perturbative approach needs additional identities 
for multi-point resolvent correlations 
even at the classical/NS limit \cite{NR_2012, E_2004}.
We have  $\cF_{\Delta}^{(1;1)}$ 
at the classical/NS limit from the result  obtained   
in \cite{CRZ_2015}:
\be
\begin{split}
\cF_{\Delta}^{(1;1)}=&1+\left[ \frac{\ell_1 \ell_{-1}}{2\hbar^2\Delta} \right]
+ \left[\frac{\ell_1^2 \ell_{-1}^2 (2 \Delta-\hbar^2)}{16\hbar^4 \Delta^3}
+\frac{(3 \ell_1^2-4\Delta \ell_2)(3 \ell_{-1}^2-4\Delta \ell_{-2})}
{16 \hbar^2 \Delta^2 (4 \Delta+3 \epsilon^2)} \right]+\cO(\eta_0^3)\,.
\end{split}
\label{11result}
\ee
where $\Delta=(c_0+\epsilon N_0)(\epsilon-(c_0+\epsilon N_0))$, 
$\ell_1=2 c_1(\epsilon-c_0)$
and $\ell_2=-c_1^2$.
$\ell_{-1}$ and $\ell_{-2}$ are its duals.
It is noted that $\Delta$ and $\ell_k$ correspond to the modes 
appearing in $\xi_2(z)$
and can be represented as the expectation values between 
regular and irregular state:
$\Delta= \langle \Delta|L_0|I^{(1)} \rangle/{\langle \Delta|I^{(1)} \rangle}$
where $\langle \Delta|I^{(1)} \rangle$ is  $Z_{(0:1)}$. 
$\ell_{+k}= \langle \Delta|L_{+k}|I^{(1)} \rangle/\langle \Delta|I^{(1)} \rangle$,
$\ell_{-k}=\langle I^{(1)}|L_{-k}|\Delta \rangle/\langle I^{(1)}|\Delta \rangle$
for   $k=1,2$
and $\langle I^{(1)}|\Delta \rangle$ is  $\bar Z_{(0:1)}$. 
This identification clearly shows that $\cF_{\Delta}^{(1;1)}$ 
satisfies the dual symmetry observed in section 2.1. 

The result \eqref{11result} is given in power of $\eta_0 =c_1c_{-1}$; 
each squared bracket in \eqref{11result} corresponds to each order of $\eta_0$.
One can easily check that the squared bracket term is consistent with the one 
appeared in \cite{BMT_201112} if one uses the series expansion in $1/\hbar^2$
at each order of $\eta_0^k$ and ignores higher order than  $1/\hbar^{2k} $.
This shows that $\cF_{\Delta}$ in powers of $1/\hbar^2$ should be considered 
as the formal series.
This fact is already suggested by Zamolodchikov and Zamolodchikov \cite{ZZ} 
for the regular conformal block
and investigated fully in \cite{RZ_2015} for the irregular conformal block:
At the classical/NS limit, 
the (ir)-regular conformal block
should behave as 
 $\cF_{\Delta} \to \exp(- F/\hbar^2)$ where the free energy $F$ is  finite as $\hbar \to 0$.

\section {Irregular conformal block with $W_3$ symmetry}
\label{sec:w3}
\subsection{Irregular matrix model and loop equation}

The irregular matrix model with $W_3$ symmetry can be  derived from  
Toda field theories at the colliding limit \cite{KMST_2013, CR_2015}. 
The simplest matrix model 
is  obtained from $A_2$ Toda theory.
The $A_2$ irregular matrix model is  two matrix model with potential  $V_1$ and $V_2$:
\be
{\mathcal Z} _{(m;n)}
=\int \prod_{i=1}^{N} \prod_{j=1}^{M} d x_i d y_j
\Delta(x)^{2 \beta} \Delta(y)^{2 \beta} \Delta(x, y)^{-\beta}
e^{\frac{\sqrt{\beta} } g \left[
\sum_{i=1}^N  V_1 (x_i)+\sum_{j=1}^M V_2 (y_j) \right]} \,,
\label{IM}
\ee
where $\Delta(x) =\prod_{i <k} (x_i -x_k)$ 
and  $\Delta(x, y) =\prod_{i , j} (x_i -y_j)$ 
are Vandermonde determinants.
$\beta$ is the deformed parameter
and conveniently put  $\sqrt{\beta} =- ib $
as in the Virasoro case. 
When $\beta=1$, the model  reduces to  hermitian two-matrix model
and the powers of the Vandermonde determinant 
correspond to the $A_2$ Dynkin index. 
The potential with rank $(m;n)$ has the explicit form  
\begin{align}
V_1(z) =
&  b_0 \log z - \sum_{k=1}^n     \frac{b_k}{k z^k} 
+ \sum_{k=1}^m     \frac{b_{-k} z^k}{k }\,,
\nn\\
V_2(z)=
& a_0  \log z - \sum_{k =1}^n    \frac{ a_k}{ k z^k}
+ \sum_{k=1}^m    \frac{a_{-k}  z^k}{ k } \,.
\label{potential-nm}
\end{align}  

To obtain the matrix model we  work  with 
the primary operator 
$ V_a(z_a) =e^{\vec \alpha_a \cdot \vec \varphi (z_a)}$
where Toda field $\vec \varphi$  
has the orthogonal  components 
$ \vec \varphi= \varphi_1 \frac{(1,1,-2)}{\sqrt 6} +\varphi_2 \frac{ (1,-1,0)}{\sqrt 2}  $
and satisfy the free field correlation 
$\v_i(z, \bar z) \v_j(w, \bar w) \sim -\delta_{i j} \log |z-w|^2$.
The conformal block with $n+m+2$  primary operators
are considered 
 with $N$ screening operators of the type $ e^{b \vec e_1 \cdot \vec \varphi (x_i )}$
and $M$  of type $ e^{b \vec e_1 \cdot \vec \varphi (x_i )}$.
Here  $\vec e_1$ and  $\vec e_2$ are  two root vectors of $A_2$. 
The conformal block is non-vanishing if the neutrality condition holds 
\cite{FL_2007}: 
\be
 \vec \alpha_\infty +\sum_a \vec \alpha_a +  bN \vec e_1 + bM  \vec e_2 = 2Q \vec \rho\,,
\ee   
where $Q=b +1/b$ is the background charge and
$\vec \rho = \vec e_1 + \vec e_2 $ is  the  Weyl vector. 

The correlation between screening operators 
provides the Vandermonde determinant with powers 
related to the Dynkin index of the two roots.
The potential is obtained if one  has the colliding limit of 
the colliding limit of the primary operators. 
If the Toda momentum of the primary operator is presented as
$ \vec \alpha_a=\alpha_a \frac{ (1,1,-2)}{\sqrt 3} +{\beta_a } (1,-1,0)  $
and $n+1$  operators are fused to the origin and $m+1$ operators at infinity,
one has  the potentials in \eqref{potential-nm} with the parameters:
$  b_k \equiv \sum_{a=0}^n   \hbar \beta_a z_a^{\, k}$ 
($k= 0, 1, \cdots, n$) 
and 
$  b_{-k} \equiv - \sum_{a=0}^n   \hbar \beta_a  z_a^{\, k}$ 
($k=   1, \cdots, m$).
Similarly, $a_\ell = (\sqrt3  c_\ell- b_\ell)/2 $ 
and $ c_k \equiv \sum_{a=0}^m \hbar \alpha_a z_a^{\, k}$
($k= 0, 1, \cdots, n$) 
and $ c_{-k} \equiv -\sum_{a=0}^m \hbar  \alpha_a z_a^{\, k}$
($k=   1, \cdots, m$). 

The duality transform  $x_i \to 1/x_i$ and 
$y_j \to 1/y_j$ induces the  dual potential  
\begin{align}
\bar V_1(z) =
&  \bar b_0 \log z - \sum_{k > 0}     \frac{\bar b_k}{k z^k} 
+ \sum_{k > 0}     \frac{\bar b_{-k} z^k}{k }\,,
\nn\\
\bar V_2(z)=
& \bar a_0  \log z - \sum_{k >0 }    \frac{ \bar a_k}{ k z^k}
+ \sum_{k >0 }    \frac{\bar a_{-k}  z^k}{ k }\,,
\label{dual-potential-nm}
\end{align} 
where $\bar a_k = - a_{-k}$ and $\bar b_k = -b_{-k}$  when $k\ne 0$.  
$\bar c_0 =c_\infty$ and $\bar a_0 =a_\infty $ are fixed by the neutrality condition 
$b_0 +  b_\infty + \epsilon (N-M/2) =\epsilon$  
and 
$a_0 + a_\infty + \epsilon (M-N/2) =\epsilon$.

The loop equation at the classical/NS limit is summarized as the following.
(More details can be found in the appendix A). 
One is the quadratic equation related with the Virasoro symmetry:
\be
X_1^2 + X_2^2 - X_1 X_2 + \epsilon (X_1' + X_2')   +   \xi_2=0\,,
\label{X-quad}
\ee 
where $X_1$ and $X_2$ are one-point resolvents 
($R_1$ and $R_2$) shifted by potential
whose integration variable is  $x_i$ and $y_i$, respectively:
$ X_1 =2\left(  R_1 + \frac13\left(2 V_1'+V_2'\right)\right)  $ and 
$X_2 =2 \left(R_2 + \frac13\left(V_1'+2V_2'\right) \right)  $.
$\xi_2(z)$ is the energy momentum tensor (Virasoro current)  expectation value 
\be
\xi_2(z) 
=-2 \epsilon (V_1'' + V_2'') 
-\frac43 (V_1'^2 + V_2'^2 + V_1' V_2')  - F
=\frac{\vev{I_m|T(z)|I_n}}{\vev{I_m|I_n}} \,. 
\label{xi_2virasoro}
\ee
where $F \equiv f_1 + f_2$ is defined in the appendix 
and has the mode expansion 
$ f_1 (z) + f_2 (z)  =\sum_{k=-{m}}^{n-1} {d_k}/{z^{k+2}} $. 
Therefore, $\xi_2(z)$ has the mode  expansion
\be
\label{su3xi2}
 \xi_2(z)=\sum_{k=-2m}^{2n} \frac{A_k}{z^{k+2}}
-\sum_{k=-{m}}^{n-1} \frac{d_k}{z^{k+2}}  \,.
\ee  
Here  $A_k$ is a constant  
\be
\label{akv}
 A_k=2 \epsilon(k+1)( a_k+b_k)
-\frac43 \sum_{r+s=k} \left( a_r a_s+  b_r  b_s +  a_r  b_s \right).
\ee
Since $ T(z)=\sum_k  {L_k}/{z^{k+2}}$,
$A_k$ ($n \le k \le 2n$)  
is the eigenvalues of $L_k$ of the irregular state  $|I_n\rangle$,
consistent with the definition of the irregular state of rank $n$.   
The mode $d_k$ has an important role since it 
is related with the partition function as in the Virasoro case.
When $ 0\le k \le n-1$, one has 
\be
d_k=  v_k \left(F_{(m;n)}  \right) \,, ~~~~
v_k =\sum_{s>0} s  \left(  b_{s+k} \fpp{ b_s}+  a_{s+k} \fpp{ a_s} \right) \,.
\label{+dk}
\ee
One may find its dual form if one replaces $b_k (a_k)$ and $k \ne 0$  
with $\bar b_k$ and $\bar a_k$, respectively. 

The other loop equation is cubic. One may put the cubic equation 
conveniently into two separate equations when 
combined with the previous quadratic one. 
\begin{align}
X_1^3+  \xi_2 X_1+ 3 \epsilon X_1 X_1'+ \epsilon^2 X_1''
&= +\frac2{3\sqrt3} \xi_3-\frac\epsilon2   \xi_2' \,,
\label{X1}
\\
X_2^3+  \xi_2 X_2+ 3 \epsilon X_2 X_2'+ \epsilon^2 X_2''
&=-\frac2{3\sqrt3}   \xi_3 -\frac\epsilon2   \xi_2' \,,
\label{X2}
\end{align}
where $\xi_3$ is given in terms of the coefficients of the potential:
\be  
  \xi_3(z)=\sum_{k=-3m }^{3n} \frac{B_k}{ z^{k+3}} 
-\sum_{k=-2m}^{2n-1}\frac{e_k}{z^{k+3}} \,.
\label{xi3-mod}
\ee
$B_k$ is  a constant 
\begin{align} 
B_k= & \frac4 {3 \sqrt{3}} \sum_{r+s+t=k} \Big(  2 (b_r b_s b_t -a_r a_s a_t) 
+ 3 (b_r b_s a_t - a_r a_s b_t ) \Big)  
\nn\\
& ~
 - \frac{\sqrt{3}}2  \epsilon  \sum_{r+s=k}
\Big( 
2(k+2 ) (b_r b_s -a_r a_s) 
+(r-s) ( b_r a_s - a_r b_s)   
\Big)
\nn\\
& ~
+\frac{\sqrt{3}}2  \epsilon^2 (k+1)(k+2) (   b_k -  a_k) \,,
\end{align}
which comes from the terms
\be
\left( 
\frac 4{3 \sqrt{3}} (2 V_1'^3+3 V_1'^2 V_2') 
+  \sqrt3 \epsilon (2 V_1' V_1'' +V_2'V_1'' )
+\frac{\epsilon^2 }3 V_1'''\right) -\Big(1 \leftrightarrow 2 \Big) \,.
\ee 
It turns out that $\xi_3(z)$ is the expectation of the $\mathcal{W}_3$ current $W(z)$
\be
\xi_3(z)=\frac{\vev{I_m|W(z)|I_n}}{\vev{I_m|I_n}}
=\sum_{k } \frac {\langle W_k \rangle}{z^{k+3}}  \,,
\label{xi_3w3current}
\ee 
and  $B_k$ ($2n \le k \le 3n$)  is the  $W_k$ eigenvalue of  the ket (right state).
When $-3n \le k \le -2n$,  $B_k$  is the  $-W_{-k}$ eigenvalue of  the bra  (left  state)
since $W_k^\dag=-W_{-k} $ (This anti-hermiticity comes from 
our normalization. See appendix B.).

The moment $e_k$ induces the flow equation. 
When $ n \le k \le 2n-1$, $e_k$ applies to the right state 
\be
e_k= \mu_k  \left(F_{(m;n)} \right) \,, ~~~~
\mu_k = \sum_{\substack{k=r+s-t;\\ t>0}}\sqrt3   {t} 
\left(( a_r  a_s +2a_r  b_s )  \frac{\partial}{\partial a_t}
- (  b_r  b_s+2a_r  b_s ) \frac{\partial}{\partial b_t} 
 \right) \,.
\label{+ek}
\ee
Its dual form applies to the left state. 

It is worth to note that if one defines  
 $\Psi_i(z)=\exp \left(\frac1\epsilon \int^z X_i(z')dz' \right)$
with $i=1,2$. 
Then, the loop equations \eqref{X1} and \eqref{X2} 
can be rewritten as a third order differential equation
of $\Psi_i(z)$: 
\be
\left(\epsilon^3 \frac{\partial^3}{\partial z^3}
+ \xi_2 \epsilon \frac{\partial}{\partial z}
+U_i(z) \right) \Psi_i(z)=0\,,
\label{3diffeq}
\ee
where $U_1(z)=+\frac2{3\sqrt3} \xi_3-\frac\epsilon2  \xi_2'$
and $U_2(z)=-\frac2{3\sqrt3} \xi_3-\frac\epsilon2 \xi_2'$.

\subsection{Spectral curve and partition function} 

As shown in sec.\  2, the symmetry present in  the spectral curve
will be used to find the partition function ${\mathcal Z} _{(m;n)}$. 
The loop equations  \eqref{X-quad}, \eqref{X1} and \eqref{X2} 
are our starting point.  
Our first step is to introduce  two  monic polynomials of $z$ with degree $N$
and $M$: 
$P (z)=\vev{\prod_{i=1}^{N} (z-x_i )} = \prod_{\alpha=1}^N (z-t_\alpha)$ 
and 
$Q (z)=\vev{\prod_j^{M} (z-y_j) }=\prod_{\alpha=1}^M (z-w_\alpha) $. 
At the classical/NS limit, one has 
the resolvents as  rational functions:
$2 R_1(z)=\epsilon P'(z) /P(z)$ and 
$2 R_2(z)=\epsilon Q'(z) /Q(z)$.  
We rewrite the quadratic equation  \eqref{X-quad} 
and cubic equations in \eqref{X1} and \eqref{X2}
in terms of the polynomials $P$ and $Q$.
The quadratic equation reduces to  the second order differential equation: 
\be
\epsilon^2 ( P'' Q  - P'  Q' + P   Q'' )
+2 \epsilon (V_1'  P'  Q  + V_2' P  Q')=F P  Q \,. 
\label{eq:quadratic} 
\ee
The cubic equations  reduces  to the third order 
differential equation:
\begin{align} 
& 
\epsilon^3 P'''+2 \epsilon^2 (2 V_1'+V_2') P''
+\epsilon \left( 4V_1'(V_1'+V_2')+2 \epsilon V_1''-F \right) P'=G_1 P
\label{cubic P}
\\
&
\epsilon^3 Q'''+2 \epsilon^2 (V_1'+2V_2') Q''
+\epsilon \left( 4V_2'(V_1'+V_2')+2 \epsilon V_2''-F \right) Q'=G_2 Q\,.
\label{cubic Q}
\end{align}
where  
\begin{align}
 G_1 = & \sum_{k=-2m}^{2n-1} \frac{1}{z^{k+3}}
\left\{
-\frac2{3\sqrt3} e_k
+ \frac23 \sum_{r+s=k}  d_r (2 b_s +a_s)   
\right\}
-\frac{\epsilon}2 \sum_{k=-m}^{n-1}  \frac{(k+2) d_k}{z^{k+3}} \,,
\\
 G_2 = & \sum_{k=-2m}^{2n-1} \frac{1}{z^{k+3}}
\left\{
+ \frac2{3\sqrt3} e_k
+ \frac23 \sum_{r+s=k}  d_r (2 a_s +b_s)   
\right\}
-\frac{\epsilon}2 \sum_{k=-m}^{n-1}  \frac{(k+2) d_k}{z^{k+3}} \,.
\end{align}

Our next step is to find the mode $d_a$ ($0 \le a \le n-1$)
and $e_k$ ($n \le k \le 2n -1$) and their duals if necessary.
As noted in section 2, it is not easy to find the exact form of $d_a$ and
and $e_k$.  
We provide some examples of the partition function 
using  the $\epsilon$ expansion method.
Note that $R_1 $ , $R_2$ and $F$ are $ O(\epsilon)$ whereas 
$V_1$ and $V_2$ are $O(1)$. Therefore, denoting 
$R_i= \sum_{k \geq 1} R_i^{(k)} \epsilon^k$,
$F= \sum_{k \geq 1} F^{(k)} \epsilon^k$
and $G_i= \sum_{k \geq 1} G_i^{(k)} \epsilon^k$,
we have $ R_1^{(1)} $ and $ R_2^{(1)}$
at the leading order of the loop equations \eqref{cubic P} and \eqref{cubic Q}:
\be
2 R_1^{(1)} =\frac{G_1^{(1)}} {4 V_1' (V_1'+V_2')} \,,~~~
2 R_2^{(1)}=\frac {G_2^{(1)}} {4 V_2' (V_1'+V_2')} \,.
\ee
The stationary point of the potentials,
$V_1'=0$, $V_2'=0$ and  $V_1'+V_2'=0$
provide the pole structure of the resolvents
(zeros of the polynomials $P$ and $Q$).   
This is the reminiscence of the cut structure on the Riemann sheet 
which appears at large $N$ limit.  
Let us denote the number of  poles 
of the resolvent  $R_1 $ by $N_k$ and $R_2$ by $M_k$
so that  $N= \sum_{k=1}^{2n} N_k$ and $M= \sum_{k=1}^{2n} M_k$.
There are equal number of stationary points for  $V_1$ and $V_2$. 
Therefore,  we have identities from the filling fractions.
When $1 \leq k \leq n$, we have 
\be 
\oint_{\cA_k}dz \, \frac{G_1^{(1)}}{4 V_1' (V_1'+V_2')}= N_k \,,
~~
\oint_{\cB_k}dz \, \frac{G_2^{(1)}}{4 V_2' (V_1'+V_2')} = M_k   \,,
\label{w3_epsilon^1_1}
\ee
where  the contours $\cA_k$ and $\cB_k$   encircle
the  stationary points of $V_1$ and $V_2$, respectively.
When $n+1 \leq k \leq 2n$, we have 
\be
\oint_{\mathcal{C}_k}dz \, \frac{G_1^{(1)}}{4 V_1' (V_1'+V_2')} = N_k \,,
~~
\oint_{\mathcal{C}_k}dz \, \frac{G_2^{(1)}}{4 V_2' (V_1'+V_2')} = M_k  \, ,
\label{w3_epsilon^1_2}
\ee
where  $\mathcal{C}_k$ encircles the stationary point of $V_1+V_2$. 
It turns out that $N_k=M_k$ since $R_1-R_2$ has no poles inside $\mathcal{C}_k$.
These identities provides $3n$-independent equations
which  solves  $d_a$ and $e_k$ in terms of the filling fraction
at the lowest order in $\epsilon$.  
One obtains the non-trivial contribution from the next order 
\be
\begin{split}
&2 R_1^{(2)}
= \frac{G_1^{(2)}}{4 V_1'(V_1'+V_2')}
+\frac{(F^{(1)}-2V_1'')G_1^{(1)}-2(2V_1'+V_2')(G_1^{(1)})'}
{16(V_1'(V_1'+V_2'))^2} 
\\
&+\frac{(2V_1'+V_2')V_1'' G_1^{(1)}}{8 (V_1')^3(V_1'+V_2')^2} 
+\frac{(2V_1'+V_2')(V_1''+V_2'') G_1^{(1)}}{8 (V_1')^2(V_1'+V_2')^3} 
-\frac{(2V_1'+V_2') (G_1^{(1)})^2}{32 (V_1'(V_1'+V_2'))^3} \,.
\end{split}
\label{w3_epsilon^2}
\ee
$R_2^{(2)}$ is obtained if $V_1$ and $V_2$ are exchanged and $G_1^{(i)} \to G_2^{(i)}$.

\subsection{Partition function ${\cal Z}_{(0;n)}$}  
 The irregular partition function ${\cal Z}_{(0;n)}$
has the potential with logarithmic and inverse powers only:
\be
 V_1(z)= b_0 \log z-\sum_{k=1}^n \frac{b_k}{k z^k} \,, ~~~
 V_2(z)=a_0  \log z-\sum_{k=1}^n  
  \frac{a_k}{ k z^k}  \,.
\label{potential(0;n)}
\ee  
This partition function is the two-point correlation 
of  one regular vertex at infinity 
and one irregular vertex at origin
and is therefore, considered as the
inner product between a regular state and an irregular state. 

The partition function is the function of $2n$-variables, 
$\{b_1, \cdots, b_n\}$ and 
$\{a_1, \cdots, a_n\}$
and  $d_a$  ($0 \le a \le n-1$)
and $e_a$ ($n \le a \le 2n-1$) 
 provide $2n$-flow equations.
In this case, $d_0$ is simple to find: 
$d_0 = 2\epsilon (b_0 N + a_0 M) 
+ \epsilon^2 (N(N-1) + M(M-1) -NM) $ 
if one uses the large $z$ expansion of the quadratic loop equation.
Other quantities need more elaborate evaluation.
  
Let us consider the partition function  ${\cal Z}_{(0;1)}$,  the rank 1 case. 
We need $e_1$. Using the results \eqref{w3_epsilon^1_1}, 
\eqref{w3_epsilon^1_1} and \eqref{w3_epsilon^2},
one finds $e_1$:
\be
\label{e1e}
\begin{split}
e_1&=2\sqrt3 \epsilon \left[ (a_1+2 b_1)(a_0 M+ b_0 N)
+3 b_0 (a_0+b_0) (N-N_1) \xi_1+3 b_0 (a_0+b_0) N_1 \xi_3 \right]
\\
&+ \sqrt3 \epsilon^2
\left[ -\frac{3 a_0\xi_3 \left( \left(N^2+ 2N_1(N_1+M+1)-N(4 N_1+M+1)\right)\xi_1
+N_1(N_1-1)\xi_3 \right)}{\xi_1-\xi_3} 
\right. \\
&\left.\quad
+ 3 b_0 \left( \left( N^2+N_1( N_1+3)-N(2 N_1+3) \right) \xi_1
+ N_1(N_1-1) \xi_3\right)+(a_1+2 b_1) d_0^{(2)} \right]
+\cO(\epsilon^3)\,,
\end{split}
\ee
where $d_0^{(2)}=N(N-1) + M(M-1) -NM$
is the $\epsilon^2$-order coefficient of $d_0$.
In addition, $\xi_1=-\frac{b_1}{b_0}$ and $\xi_3=-\frac{a_1+b_1}{a_0+b_0}$  
are stationary points of $V_1$ and $V_1+V_2$, respectively.
Using the flow equations $d_0=v_0 \left( F_{(0:1)} \right)$
and $e_1= \mu_1 \left( F_{(0;1)} \right)$, one can find the free energy
and  the partition function up to  $O(\epsilon^2)$
\be
\begin{split}
\cZ_{(0;1)}=&a_1^{-(2 \epsilon a_0 M_1
+\epsilon^2 M_1 (M_1-1))/\hbar^2}
b_1^{-(2 \epsilon b_0 N_1+\epsilon^2 a_0 N_1(N_1-1))/\hbar^2}
(a_1+b_1)^{-(2 \epsilon N_2 (a_0+b_0)+\epsilon^2 N_2(N_2-3))/\hbar^2} 
\\
&\times (a_0 b_1-a_1 b_0)^{-\epsilon^2(N_1 N_2+ N_2+ M_1 N_2-M_1 N_1)/\hbar^2}\,.
\end{split}
\ee

\subsection{Partition function ${\cal Z}_{(m;n)}$ and ICB}  

As in the Virasoro case, ${\cal Z}_{(m;n)}$ is equivalent to 
the two-point correlation of  irregular vertex operators. 
One can evaluate 
the irregular conformal block (ICB) using 
the relation  with the partition function \cite{CR_2015}
\be
{\cal F}_\Delta^{(m;n)}  (\{a_{-k }, b_{-k }  : a_k, b_k \})= 
\frac{e^{\zeta_{(m;n)}}   
{\cal Z}_{(m;n)}  (a_0, b_0 ; \{a_{\ell },  b_{\ell}  \} )} 
{\cZ_{(0;n)}  (a_0, b_0 ; \{a_k, b_k \} )  \cZ_{(0;m)}   
(\bar a_0, \bar b_0 ; \{\bar a_k, \bar b_k \} )) }\,. 
\label{w3:ICB}
\ee   
This time the  extra factor $e^{\zeta_{(m;n)}}$
is the generalization of the Virasoro case:  
$\hbar^2 \zeta_{(m:n)} =-\sum_k^{{\rm min}(m,n)} \frac{4}{3k} 
(2 a_k a_{-k}+a_k b_{-k}+b_k a_{-k} +2 b_k b_{-k})$.

We find the partition function ${\cal Z}_{(1;1)} $
using the  $\epsilon$ expansion. 
By rescaling $x_i \to x_i/ a_{-1}$ and $y_j \to y_j/a_{-1}$,
one obtains $a_{-1}$ dependence and has
the partition function with three parameters,
$\eta_0\equiv a_1 a_{-1}$, $t_1 \equiv b_1/a_1$
and $t_{-1}\equiv b_{-1}/a_{-1}$.
Then, the partition function is to be evaluated
from three flow equations:
$d_0=v_0(F_{(1;1)})$, $e_1=\mu_1 (F_{(1;1)})$
and $e_{-1}=\mu_{-1} (F_{(1;1)})+\nu_{-1}^c$
where $\nu_{-1}^c$ is a constant (See appendix A).
We use notations for the filling fractions as
$M_0=M_1+M_2$, $N_0=N_1+N_2$ $(M_2=N_2)$
and $M_\infty=M_{-1}+M_{-2}$, 
$N_\infty=N_{-1}+N_{-2}$ $(M_{-2}=N_{-2})$
with $M=M_0+M_\infty$, $N=N_0+N_\infty$.
We have up to $\cO(\epsilon^1)$,
\be
\label{d011}
d_0=2 \epsilon( a_0 M_0+ b_0 N_0)+\cO(\eta_0)\,,
\ee
\be
\label{e111}
e_1=
2\sqrt3 \epsilon \big[a_0 \left(M_0 a_1+(2M_0-3 N_0+3N_1) b_1\right)
+b_0 \left( (N_0-3N_1) a_1-N_0 b_1 \right) \big]
+\cO(\eta_0)\,,
\ee
\be
\label{e-111}
\begin{split}
e_{-1}-\nu_{-1}^c &=
-2\sqrt3 \epsilon \big[a_0 \left(M_\infty a_{-1}+(2M_\infty-3 N_\infty
+3N_{-1}) b_{-1}\right)
\\
& \qquad \qquad ~~~
+b_0 \left( (N_\infty-3 N_{-1}) a_{-1}-N_\infty b_{-1} \right) \big]
+\cO(\eta_0)\,,
\end{split}
\ee
and the partition function
\be
\label{z11}
\begin{split}
\cZ_{(1;1)}&=a_1^{-(2 \epsilon a_0 M_1)/\hbar^2}
a_{-1}^{-(2 \epsilon a_\infty M_{-1})/\hbar^2}
b_1^{-(2 \epsilon b_0 N_1)/\hbar^2}
b_{-1}^{-(2 \epsilon b_\infty N_{-1})/\hbar^2}
\\
&\times 
(a_1+b_1)^{-2 \epsilon (a_0+b_0)/\hbar^2}
(a_{-1}+b_{-1})^{-2 \epsilon (a_\infty+b_\infty)/\hbar^2}
\\
&\times  \mathrm{exp} \bigg[
\frac{ 2 \epsilon \eta_0}{\hbar^2 a_0 b_0(a_0+b_0)}
\Big\{ b_0^2 (M_1-M_{-1})+a_0^2 (N_1-N_{-1})t_1 t_{-1}
+a_0 b_0\big( M_1-M_{-1}
\\
& \qquad ~~~  +(N_2-N_{-2})(1+t_1+t_{-1})
+(N_1+N_2-N_{-1}-N_{-2}) t_1 t_{-1} \big)
\Big\}+\cO(\eta_0^2) \bigg]\,.
\end{split}
\ee
This provides ICB
\be
\begin{split}
\cF_\Delta^{(1;1)}=&\left[ 1-\frac{4\eta_0}{3\hbar^2}
(2+ t_1+t_{-1}+2t_1 t_{-1})+\cO(\eta_0^2) \right]
+\epsilon \bigg[ \frac{2\eta_0}{\hbar^2 a_0 b_0(a_0+b_0)}
\Big\{ b_0^2 (M_1-M_{-1})
\\
&+a_0^2 (N_1-N_{-1})t_1 t_{-1}
+a_0 b_0\big( M_1-M_{-1}+(N_2-N_{-2})(1+t_1+t_{-1})
\\
&+(N_1+N_2-N_{-1}-N_{-2}) t_1 t_{-1}\big) \Big\} 
+\cO(\eta_0^2) \bigg]+\cO(\epsilon^2) \,.
\end{split}
\label{W3_(1:1)ICB}
\ee

One may use the perturbative method to find ICB
using IMM with the relation \eqref{w3:ICB}. 
One may put  the reference potentials $V^{(0)}$ and
its perturbations $\Delta V^{(0)}$
for $N_0$ and $M_0$ variables: 
\be
\begin{split}
V^{(0)}(x_I, y_J)
&= \sum_{I=1}^{N_0} \Big(b_0 \log x_I - 
 \sum_{k=1}^{n} \frac{b_k} {k  x_I^{k}} \Big)
 +\sum_{J=1}^{M_0} \Big(a_0 \log y_J - 
 \sum_{k=1}^{m} \frac{a_k} {k  y_J^{k}} \Big)
\,, \\
 \Delta V^{(0)}(x_I,y_J) & =
  \sum_{I=1}^{N_0} \Big(
  \sum_{\ell=1}^{n} \frac { b_{-\ell}}{\ell}  { x_I^{\ell}} \Big)
  + \sum_{J=1}^{M_0} \Big(
  \sum_{\ell=1}^{n} \frac { a_{-\ell}}{\ell}  { y_J^{\ell}} \Big) \,.
 \end{split} 
\ee
For the rest variables,  $ N_\infty$ and $M_\infty$ variables,  
one has the reference potential 
$V^{(\infty)}(\mu_K, \nu_L)$ and perturbation $\Delta V^{(\infty)}(\mu_K, \nu_L)$
which can be put into the similar form $V^{(0)}$, $\Delta V^{(0)}$
with dual variables if one uses the dual transformation 
 $\mu_K \to 1/\mu_K$ and $\nu_L \to 1/\nu_L$. 
After this, one has ICB in the following form
\be
\begin{split}
{\cal F}_\Delta^{(m;n)}  (\{a_{-k }, b_{-k }  : a_k, b_k \})= 
e^{\zeta_{(m;n)}} 
&\Big \langle 
\prod_{I, K}  (1 -x_I \mu_K)^{2\beta}  
\prod_{J, L}  (1 -y_J \nu_L)^{2\beta}  
\prod_{I, L}  (1 -x_I \nu_L)^{-\beta}  
\\
&\times
\prod_{J, K}  (1 -y_J \mu_K)^{2\beta}  
 e^{ \frac{\sqrt{\beta}}g 
\left (\Delta V^{(0)} (x_I,y_J) + \Delta V^{(\infty)} (\mu_K, \nu_L) \right) }
\Big \rangle_{\!\! 0}\,,
\end{split}
\ee
where the bracket denotes the expectation value with respect to
the reference potentials, $V^{(0)}$ and $V^{(\infty)}$.
One may obtain the expectation values 
using the large $z$ expansion of the resolvents 
in the loop equations 
\eqref{X1} and \eqref{X2}
of the reference partition functions.

We find ${\cal F}_\Delta^{(1;1)}$ as the simplest example.   
Up to the first order of $a_1$ and $b_1$ (also their duals $\bar a_1$ and  $\bar b_1$), we have 
\be
\begin{split}
\cF_\Delta^{(1;1)}&=1+\frac1{\hbar^2} \Big[
2 \epsilon^2 \left( \vev{x_I} \vev{\mu_K}+\vev{y_J} \vev{\nu_L} \right)
-\epsilon^2 \left( \vev{x_I} \vev{ \nu_L}+ \vev{ y_J} \vev{\mu_K} \right)
\\
&+2 \epsilon \left( \bar b_1 \vev{x_I}+\bar a_1 \vev{y_J}
+b_1 \vev{\mu_K}+a_1 \vev{\nu_L} \right)
+ \frac{4}{3} 
(2 a_1 \bar a_1+a_1 \bar b_1+b_1 \bar a_1 +2 b_1 \bar b_1)\Big] \,.
\end{split}
\ee
Here we omitted summation symbols
inside the expectation value bracket for simplicity.
Each expectation values can be read off from
the order of $z^{-4}$ of the loop equations \eqref{X1} and \eqref{X2} for
the reference partition function $\cZ_{(0;1)}$.
Finally, we obtain ICB at the first order of $a_1$ and  $b_1$
\be
\begin{split}
\cF_\Delta^{(1;1)}=
1-&\frac1{9 \hbar^2 \left(4 \omega_0^2+\Delta^2 (4 \Delta-3 \epsilon^2) \right)}
\bigg[8 \Delta \omega_{-1} \omega_1
\\
&\qquad \qquad  +12 \omega_0( \omega_{-1} \ell_1
+\omega_1 \ell_{-1} )
-\frac92 \Delta \ell_{-1} \ell_1
(4 \Delta-3 \epsilon^2 ) \bigg]\,,
\end{split}
\label{W3-ICB11}
\ee
where $\Delta=-\frac43(\alpha^2+\alpha \beta
+\beta^2)+2\epsilon(\alpha+\beta)$
with $\alpha=a_0+\epsilon(M_0-N_0/2)$, $\beta=b_0+\epsilon(N_0-M_0/2)$,
$\ell_{1} =-\frac43 (a_0(2a_1+b_1)+b_0(a_1+2b_1))+4\epsilon(a_1+b_1)$
and its dual $\ell_{-1}$ are the constant modes of $\xi_2(z)$.
As in the Virasoro case, we may identify the expectation values of 
the Virasoro generators: 
$\ell_{1}= \frac{\langle \Delta|L_{1}|I^{(1)} \rangle}{\langle \Delta|I^{(1)} \rangle}$, 
$\ell_{-1}=\frac{\langle I^{(1)}|L_{-1}|\Delta \rangle}{\langle I^{(1)}|\Delta \rangle}$
and 
$\Delta= \frac{\langle \Delta|L_0|I^{(1)} \rangle}{\langle \Delta|I^{(1)} \rangle} $
where ${\langle \Delta|I^{(1)} \rangle}$ is $\cZ_{(0;1)}$
and ${\langle I^{(1)}|\Delta \rangle}$ is $\bar{\cZ}_{(0;1)}$. 
In addition, $\omega_k$ is the mode appearing in  $\xi_3(z)$.
The constant mode $\omega_0=\frac1{3\sqrt3}
(\alpha-\beta)(4\alpha+2\beta-3\epsilon)(2\alpha+4\beta-3\epsilon)$ is  identified as  
$ \frac{\langle \Delta|W_0|I^{(1)} \rangle}{\langle \Delta|I^{(1)} \rangle} $  
and the other modes are expectation values: 
$\omega_1= \frac{\langle \Delta|W_1|I^{(1)} \rangle}{\langle \Delta|I^{(1)} \rangle} 
=B_1-e_1$
and
$\omega_{-1}$ is its dual.   
We check that the $\epsilon$ expansion of  the above $ \cF_\Delta^{(1;1)}$ in 
\eqref{W3-ICB11} is
in complete agreement with \eqref{W3_(1:1)ICB}. 
It is also noted that ICB is manifestly dual invariant.

ICB of \eqref{W3-ICB11}, obtained from 
the perturbation of the irregular matrix model 
is convenient to find the irregular state in terms of descendants. 
The irregular state of the  rank 1 has the form  
\be
\begin{split}
 \frac{|I^{(1)} \rangle}{\langle \Delta|I^{(1)} \rangle} 
=1-&\frac1{9 \hbar \left(\omega_0^2+\Delta^2(\Delta-\frac{c-2}{32} \right)}
\Bigg[(2 \Delta \omega_1-3\omega_0 \ell_1 ) W_{-1} |\Delta \rangle
\\
&\qquad 
\qquad -\left(3 \omega_0 \omega_1+\frac92 \Delta \left(\Delta-\frac{c-2}{32} 
 \right)\ell_1\right)
L_{-1} |\Delta \rangle \Bigg]+ \cdots  
\end{split}
\label{W3:irregularstate1}
\ee 
where $c=2+24\epsilon^2$ is the central charge and $\cdots$ refers to the higher descendant. 
The irregular state \eqref{W3:irregularstate1}
has no semi-degenerate condition at the first level 
in contrast to the state constructed in \cite{KMST_2013, W_2009, KMST_2009}
where $L_{-1}$ is related to $W_{-1}$ descendant. 
Instead, the coefficient $\omega_1$ is not a simple constant 
and is given in terms of the flow equation 
with respect to the proper normalization $\cZ_{(0:1)}$.
This feature also appeared in Virasoro irregular state 
with rank 2 and higher \cite{CRZ_2015}. 
However,  here in Toda irregular state, the non-trivial 
 feature appears even for the rank 1 and at the first descendant level. 
 
\section{Relation with gauge theories}
\label{sec:gauge}

The Nekrasov partition function  for $\mathcal{N}=2$ 
is given  as the product 
of the perturbative part and instanton part:
\be
 Z_{\mathrm{full}}(\vec q;\vec a, m;\epsilon)= Z_{\mathrm{perturb}}
Z_{\mathrm{inst}}= Z_{\mathrm{tree}}
Z_{\mathrm{1loop}}Z_{\mathrm{inst}}\,.
\ee 
According to AGT, the Nekrasov partition function 
of $U(N)$ gauge theory with $N_f=2N$  is 
related with the correlation function of Toda field theory. 
Explicitly, the correlation function on a sphere with $n + 3$ punctures 
corresponds to a linear quiver $n$ of $SU(N)$ gauge groups:
\ba
\langle V_n(\infty)V_{n-1}(1) V_{n-2}(q_1)\cdots
V_2(q_1\cdots q_{n-3}) V_1(0)\rangle
\propto \int | Z_{\mathrm{full}}(\vec q;\vec a, m;\epsilon)|^2
\ea
Especially the instanton part of Nekrasov partition function 
$Z_{\mathrm{inst}}$ corresponds to the conformal blocks ${\mathcal F_\Delta}$ 
in the CFT side; 
 $ Z_{\mathrm{inst}}={\mathcal F_\Delta}$. 
One may  try to prove this conjecture by evaluating the conformal block directly 
\cite{IO_2010, Mironov:2010pi, Zhang:2011au, AFLT_2010}. 

One may equally check the AGT conjecture 
by comparing the Seiberg-Witten curve of the gauge theory with the 
corresponding  spectral curve from CFT side. 
Seiberg-Witten curve of $SU(N)$ gauge theory with $N_f=2N$ is given in 
\cite{IMT_2009, Witten, G_2009a},
\be
x^n 
= \sum_{k=2}^n \frac{P_{2k}^{(k)}(z)}{ (z(z-1)(z-q ) ))^k}
x^{n-k}\,,
\ee
where $P_{2k}(z)$ is a $2k$ polynomial in $z$
and $q=e^{i \pi \tau}$  with  $\tau$  
the gauge coupling constant $\tau= \theta/\pi + 8\pi i/g^2$. 

On the other hand, from the matrix model with $A_{n-1}$  quiver 
one may have the spectral curve of the type 
\be
x^n = \sum_{k=2}^{n} (-1)^{k-1} \phi_k(z) ~ x^{n-k}, 
\ee
where $ \phi_k(z)  $ is the expectation value of the conformal current with  dimension $k$. 
This was obtained from the the expectation value of the 
Miura transform of $ \det (x - i \hbar \partial \varphi (z)) =0 $ 
with Toda field $\varphi$ in the matrix model context. 
Even though this spectral curve holds at $\epsilon \to 0$ limit 
(Note that this limit corresponds to the large $N$ limit where $N$ is the number of the eigenvalues of the matrix), 
one may  identify  parameters between two theories  
so that the pole structure of the curves matches each other as done in \cite{IMT_2009}.
One may have more accurate spectral curve if one uses the 
loop equation of the matrix model. 

This conjecture still works for irregular conformal blocks.
Gaiotto \cite{G_2009} obtained the irregular singularity 
for the simplest example of $\mathcal W_2$ 
and identify the result with the asymptotically free theories of $SU(2)$ gauge theory 
with $N_f <4$.
Explicitly, $SU(2)$ gauge theory with  $N_f=4$  reduces to $ SU(2)$ 
pure Yang-Mills theory
if the masses $\mu_i$ of  hyper-multiplets  are to infinity 
$\mu_1,\ldots,\mu_4 \rightarrow \infty$
and $q\rightarrow 0$, while keeping 
$
q\prod_{I=1}^4 \mu_I = \Lambda^4
$. 
Then the Seiberg-Witten curve is given as $x^2 =\phi_2(z)$  where 
\ba
\phi_2=\frac{\L^2}{z^3}+\frac{2u}{z^2}+\frac{\L^2}{z}\,,
\ea
which contains the strongest irregular singularity with odd power on the Riemann sphere. 
$u$ parameterizes the Coulomb branch.
This irregular singularity of odd power is not properly represented in terms of matrix model as yet
unless one considers the special limit from the even power singularity case
so that the highest even power term vanishes.

For $N_f=2$, one may have the SW curve  with 
the  irregular singularity with degree 4
\ba
\phi_2=\frac{\L^2}{z^4}+\frac{2m\L}{z^3}+\frac{2u}{z^2}+\frac{2\tilde m\L}{z}+\L^2
\ea
where 
$
q \mu_2\mu_4 =4 \Lambda^2
$, $m=\mu_1-\frac{\epsilon}{2}$ and $\tilde m=\mu_3-\frac{\epsilon}{2}$.  
In this case $\phi_2$ is   $\xi_2$ in  \eqref{eq:xi2}  with $n=m=1$,  and 
the Nekrasov instanton partition function is checked to be the same as the partition function 
${Z}_{(1;1)}$ of the irregular matrix model of the rank (1;1)  \cite{EM_2010, NR_2011}. 
In addition, ${Z}_{(1;1)}$ corresponds to the inner product of two irregular states
whose eigenvalues of $L_2$ and $L_1$ are given from $\phi_2$;
$\L^2$ and $2m\L$ for the ket (irregular module) and 
$\L^2$ and $2\tilde m\L$ for the bra,  respectively.
The Coulomb branch parameter $u$ is related to $d_0$
and is fixed by the filling fraction of the matrix model, the contour integral of the resolvent,
which is identified with the contour integral of the Seiberg-Witten one-form $x dz$. 
In addition, from  the fixed $d_0$ the partition function is obtained from the corresponding 
Virasoro flow equation as seen in section 2. 
Considering this result, 
one may naturally assume that the AGT conjecture works for the colliding limit as well  
with higher ranks $(m;n)$. 
 For the special case  $(0;n)$, one has the irregular singularity of type $D_{2n}$ considered 
in \cite{CV_2011, BMT_201112}
which is  associated with a regular puncture of degree 2 and an irregular puncture 
of degree $ 2n + 2$  on the  sphere.  
Specifically, ${Z}_{(0;1)} $ corresponds to $D_2$ and  ${Z}_{(0;2)} $ 
corresponds to $D_4$. 

One may extend the analysis to the case with $\mathcal {W}_N$ symmetry. 
For $\mathcal W_3$, one has the SW curve with the cubic equation. 
One can have the same spectral curve from the Penner-type matrix model
with the multi-log potential
\be
V_1(z)=\sum_{k=1}^p m_k \log (q_k-z) \,,
~~~~
V_2(z)=\sum_{k=1}^p \tilde{m}_k \log (q_k-z) \,,
\label{multi-log}
\ee
which corresponds to the linear quiver  $p-2$  $SU(3)$ gauge theories \cite{G_2009a}.
For instance, in \cite{IMT_2009}, the potential \eqref{multi-log} for $p=3$
was considered,
which corresponds to $SU(3)$ gauge theory with six massive flavors.
The relation between six mass parameters and parameters in the potential
was made by using the  residue relation  of the one-form $x dz$.

At the colliding limit, we used the potential 
\eqref{potential-nm}
and obtained the two cubic equations \eqref{X1} and \eqref{X2}.
In fact, the  two cubic equations reduce to one in the large $N$ limit $(\epsilon \to 0)$,
\be
x^3+  \frac{\xi_2}4 x-\frac{\xi_3}{12\sqrt3} =0 \,,
\label{spectralcurve}
\ee
where $x=2 X_1$ or $-2 X_2$.
Note that the spectral curve \eqref{spectralcurve} has the same form  
even before the colliding limit, where the explicit form of $\xi_2$, $\xi_3$ 
is different. 
At the colliding limit, 
$\xi_2$ and $\xi_3$ have irregular singularities
which indicate positive modes of the Virasoro and $\mathcal{W}_3$ currents
as we have shown in \eqref{xi_2virasoro} and \eqref{xi_3w3current}. 
In addition, the spectral curve provides further information about the gauge theory 
with the same SW curve.
For example, if one considers the potential 
\eqref{potential(0;n)} of the rank $(0;n)$,
then the spectral curve (at $\epsilon \to 0$ limit) 
is identified with the SW curve of the type IV Argyres-Douglas theory 
\cite{X_2012}:
\begin{align}
x^3&+\left(\frac{v_1}{z^{2n+2}}+\frac{v_2}{z^{2n+1}}+\cdots+\frac{v_{2n}}{z^{3}}
+\frac{v_{2n+1}}{z^2} \right)x  \nn \\
& +\left(\frac1{z^{3n+3}}+\frac{\omega}{z^{3n+2}}
+\frac{u_1}{z^{3n+1}}+\cdots+\frac{u_{3n-2}}{z^4}+\frac{u_{3n-1}}{z^3} \right)
=0\,,
\end{align}
where the dominant singular part in $\xi_3$ is normalized as 1. 
Comparing this SW curve  with the explicit expression $\xi_2$ in  \eqref{su3xi2} and 
$\xi_3$ in \eqref{xi3-mod},
one notes that $(2n+1)$ parameters 
($v_1, \cdots, v_{n+1}$ and $\omega, u_1, \cdots, u_{n-1}$)
are fixed by the $(2n+2)$  parameters of the potential \eqref{potential(0;n)}
with the proper normalization.  
The mass parameters $v_{2n+1}$ and $u_{3n-1}$ are determined by
the additional constants $d_0$ and $e_0$. 

Finally, the Coulomb branch parameters $v_{n+2}, \cdots, v_{2n}$
and $u_n, \cdots, u_{3n-2}$
are related with the contour integral of the SW one-form in the gauge theory side. 
From the matrix model side, the Coulomb branch parameters 
are given in terms of  $d_i$ $(i=1, \cdots, n-1)$ and $e_j$
$(j=1, \cdots, 2n-1)$ 
and can be fixed by $(3n-2)$ independent filling fractions
which is related with the cut structure of the spectral curve. 
Once the Coulomb branch parameter are known,
one can find the partition function
since the values $d_k$ and $e_k$ 
are directly related with the flow equations,
which are the strong merit of the matrix model approach.   


\section{Conclusion and outlook}
\label{sec:summary}

We develop  a new mechanism to evaluate the irregular conformal block 
using the Virasoro and W symmetry. 
We use the loop equation of the irregular matrix model
which encodes all the details of the conformal symmetry. 
At the classical/NS limit, the loop equation does not contain 
the multi-point resolvent terms and reduces to the simple spectral curve
which contains the first derivative of the resolvent. 
The special feature of the spectral curve is that 
it contains not only constants of motion but also flow equations
corresponding to the conformal symmetry.
The flow equations are defined on the parameter space of the potential  of 
the irregular matrix model,
and its generators represent the Virasoro and W symmetry. 
We present the details of the flow equations 
and how to obtain the partition function and irregular conformal block. 
The irregular conformal block 
is related with the partition function of the  Argyres-Douglas theory
according to AGT conjecture,
if one uses the  parameter relations between these two theories 
whose details can be found, for example, in \cite{KMST_2013}. 
    
It is noted that the spectral curve and flow equation are not restricted to
the irregular conformal block. 
The method can be applied to the regular conformal block at the classical/NS limit.
Using the similar flow equation, one can find the partition function 
\cite{NR_2012, RZ_2015a}.
Even though the partition function is simply obtained, 
the relation of the  positions of the primary operators is not.
For example, 5-point Liouville conformal block 
with one degenerate operator 
reduces to Painlev\'e VI as  presented in \cite{LLNZ_2013}. 
It seems to be worthwhile to investigate the connection between the positions of 
the multi-point regular conformal block.  
 
Nekrasov partition function and its counter part, 
regular conformal block are represented in terms of Young diagrams 
\cite{Nekrasov1, Nekrasov2, AFLT_2010}.
Irregular conformal block should also be represented in the same way, 
which is not well understood yet. 
On the other hand, conformal symmetry is reinstated in the \emph{degenerate
double affine Hecke algebra}(DDAHA)  
and Nekrasov partition function was studied in terms of DDAHA \cite{KMZ_2013}. 
In the same way, the irregular conformal block can be better
understood using DDAHA. 
There was a few attempts to investigate this connection \cite{MRZ_2014, Bourgine_2014} 
and it should be worth finding  DDAHA representation 
of the irregular conformal block.  

Finally, the mixture of bulk and micro Coulomb charges 
in two dimensions is an interesting system whose
interaction is represented in terms of the logarithmic potential. 
If the system is fine-tuned so that the system  
shows the conformal symmetry, then the matrix model should 
play the role. In addition, if the bulk charges are localized so that 
they are idealized in terms of finite number of multi-poles, 
then the free energy of the irregular matrix model can be useful.  

\subsection*{Acknowledgements}
This work is supported by the National Research Foundation of Korea(NRF) grant funded by the Korea government(MSIP) (NRF-2014R1A2A2A01004951).

\appendix

\section{Loop equation of $A_2$ irregular matrix model}
\label{app:loop}
The $A_2$ irregular matrix model \eqref{IM} has the Virasoro and $W_3$ symmetry 
which is represented in terms of  
loop equations \cite{CR_2015, KLLR_2003, NSW_2003, SW_2009}.  
We put the  multi-point resolvent   as
\be
R_{K_1;\cdots ;K_s}(z_1,\cdots, z_s)
=\beta\left(\frac g{\sqrt\beta} \right)^{2-s}
\vev{\sum_{i_1=1}^{N_{K_1}} \frac1{z_1-\l_{i_1;K_1}}
\cdots \sum_{i_s=1}^{N_{K_s}} \frac1{z_s-\l_{i_s;K_s}}}_{\mathrm{connected}} \,.
\ee
where denoting $\l_{i;1}=x_i$, $\l_{j;2}=y_j$. One obtains
the quadratic loop equation if one performs the conformal transformation 
of the integration variables $x_i  \to x_i+ \ep/(x_i-z)$
and $y_j  \to y_j + \ep/(y_j -z)$
which provides the Virasoro symmetry:
\be
\begin{split}
&R_1(z)^2 +R_2(z)^2- R_1(z) R_2(z)+V_1'(z) R_1(z)+V_2'(z) R_2(z)\\
&+\frac{\hbar Q}2 \left(R_1'(z)+R_2'(z)\right)-\frac{\hbar^2}4 \left(
R_{1;1}(z,z)-R_{1;2}(z,z)+R_{2;2}(z,z) \right)
 = \frac{f_1(z)+f_2(z)}4\,,
 \end{split}
\ee
where  $f_1(z):=4 g \sqrt{\beta} \sum_i^{N_1} 
\vev{  \frac{V_1'(z)-V_1'(x_i)}{z-x_i}}$
and 
$f_2(z):=4 g \sqrt{\beta} \sum_i^{N_2} 
\vev{  \frac{V_2'(z)-V_2'(y_j )}{z-y_j }}$.
Here $\vev{\cdots}$ denotes the expectation value 
with respect to the $A_2$ matrix model.

$W_3$ symmetry is given in terms of cubic loop equation \cite{CR_2015}
\be
\begin{split}
&0=-R_1^2 R_2+R_1 R_2^2-V_1'(R_1^2+V_1' R_1-\frac{f_1}4)
+V_2'(R_2^2+V_2' R_2-\frac{f_2}4)+\frac{g_1-g_2}4\\
&+\frac{\hbar Q}4
\left[ 3(V_2' R_2'-V_1' R_1')+R_1 R_2'-R_1' R_2+2(R_2 R_2'-R_1 R_1')
+V_2'' R_2-V_1'' R_1+\frac{f_1'-f_2'}4 \right] \\
&+\frac{\hbar^2 Q^2}8(R_2''-R_1'')
+\frac{\hbar^2}4\left[V_1' R_{1;1}-V_2' R_{2;2}+
R_{1;1} R_2-R_{2;2} R_1-2 R_{1;2}(R_2-R_1) \right] \\
&+\frac{ \hbar^3 Q}{16} \left[R_{1;1}'-R_{2;2}'
+\lim_{\bar z \to z} \left( \fpp{z} R_{1;2} (z , \bar z)
-\fpp{\bar z} R_{1;2}(z, \bar z) \right) \right]
+\frac{\hbar^4}{16}(R_{1;2;2}-R_{1;1;2})\,,
\end{split}
\ee
where 
$g_1(z) := 4 g^2 \beta \sum_{i,j}
\vev{\frac{V_1'(z)-V_1'(x_i)}{(z-x_i)(x_i-y_j)}}$
and $g_2(z) := 4 g^2 \beta \sum_{i,j}
\vev{\frac{V_1'(z)-V_2'(y_j)}{(z-y_j)(y_j-x_i)}}$.
This is obtained after varying  the integration variables
$x_i \to x_i+\sum_{j=1}^{N_2} \frac{\epsilon}{(x_i-z)(x_i-y_j)}$
and
$y_j \to y_j+\sum_{i=1}^{N_1} \frac{\epsilon}{(y_j-z)(x_i-y_j)}$.

At the classical/NS limit 
($\hbar \to 0$, $b \to \infty$ while $\hbar b = \epsilon $ finite),
each multi-point resolvent is finite but due to the factor $\hbar$,
the multi-point resolvent terms drop out 
and the loop equations are given in the simple form:
\begin{gather}
X_1^2+X_2^2-X_1 X_2+2 \epsilon (X_1'+X_2')=-\xi_2\,,
\\
X_1^2 X_2-X_1 X_2^2+\frac\epsilon2
\left[(2X_1+X_2)X_1'-(X_1+2X_2)X_2' \right] 
+\frac{\epsilon^2}2 (X_1''-X_2'')=\frac2{3\sqrt3} \xi_3\,,
\end{gather}
where $ X_1/2 = R_1 +(2 V_1'+V_2')/3   $,
$X_2/2 =R_2 + (V_1'+2V_2')/3  $ and
\begin{align}
\xi_2=&-2 \epsilon (V_1''+V_2'')
-\frac43\left[(V_1')^2+V_1' \, V_2'+(V_2')^2 \right]
-(f_1+f_2)\,,
\\
\begin{split}
 \xi_3=
&
\frac{4\sqrt3}{9} 
\left( 2 (V_1')^3+3 (V_1')^2 V_2'-3 V_1' (V_2')^2-2(V_2')^3 \right) \\
&+\sqrt3 \epsilon
(2 V_1' V_1''+V_2' V_1''-2V_2' V_2''-V_1' V_2'')
+\frac{\sqrt3}{2} \epsilon^2 (V_1'''-V_2''') \\
&+ \sqrt3 \left((f_1-2f_2)V_1'+(2 f_1-f_2)V_2' \right)
+3 \sqrt3 (g_1-g_2)+\frac{3\sqrt3}4 \epsilon (f_1'-f_2')\,,
\end{split}
\end{align}

$\xi_2$ and $\xi_3$ look complicated but can be written in a compact form 
if one uses the mode expansion,
\be
 \xi_2=\sum_{k=-2m}^{2n} \frac{A_k}{z^{k+2}}
-\sum_{k=-m}^{n-1} \frac{d_k}{z^{k+2}}
\,, ~~~~~~
 \xi_3=\sum_{k=-3m }^{3n} \frac{ M_k}{ z^{k+3}} 
-\sum_{k=-2m}^{2n-1}\frac{e_k}{z^{k+3}} \,,
\label{app_xi}
\ee
where $A_k$ and $B_k$ are constants (here we use the notation
$c_\ell$ which is related with $a_\ell=(\sqrt3 c_\ell-b_\ell)/2$ )
\begin{align}
A_k=&\epsilon (k+1)(b_k +\sqrt3  c_k)
-\sum_{r+s=k} \left( b_r  b_s 
+ c_r c_s \right)\,, \nn \\
\begin{split}
B_k= &\sum_{r+s+t=k}( 3c_r \, b_s \, b_t-c_r \, c_s\, c_t) 
 +\frac32 \epsilon \sum_{r+s=k} (s+1)(\sqrt3 c_r c_s-2 c_r b_s-\sqrt3 b_r b_s) \\
& +\frac34 \epsilon^2 (k+1)(k+2) (\sqrt3 b_k-c_k) \,.\nn
\end{split} 
\end{align}
$d_k$ is the mode of $f_1+f_2= \sum_{k=-{m}}^{n-1} \frac{d_k}{z^{k+2}}$
and induces the flow equation 
when $-(m-1) \le k \le n-1$;
\be
d_k = \left\{ \begin{array}{ll}
 v_k^\partial \, \left( -\hbar^2 \log\cZ_{(m;n)} \right) \,,  &0 \leq k \leq n-1\\
 \\
2 \epsilon (b_k N_b+c_k N_c)+
u_k^\partial  \, \left(-\hbar^2 \log\cZ_{(m;n)} \right)
\,, & -(m-1) \leq k \leq -1 
\end{array} \right.
\label{app:d_k rep}
\ee
\be
v_k^\partial=\sum_{\substack{r-s=k \\ 0 < s \\}} s
\left(  b_r \fpp{ b_s}+ c_r \fpp{ c_s} \right) \nn
 \,, ~~~~
u_k^\partial=\sum_{\substack{r-s=k \\  s<0 \\}} (-s)
\left(  b_r \fpp{ b_s}+ c_r \fpp{ c_s} \right) \nn
\ee
where $N_b=N-M/2$, $N_c = \sqrt3 M/2$. 
On the other hand, $d_{-m}$ is constant, 
$d_{-m}=2 \epsilon (b_{-m} N_b+c_{-m} N_c)$.
 
The mode  $e_k$ is defined as
\be
-\sum_{k=-2m}^{2n-1}\frac{e_k}{z^{k+3}} 
=  \sqrt3 \left((f_1-2f_2)V_1'+(2 f_1-f_2)V_2' \right)
+3 \sqrt3 (g_1-g_2)+\frac{3\sqrt3}4 \epsilon (f_1'-f_2') \nn\,,
\ee
and also induces the flow equation. 
When $ -(2m-1) \le k \le 2n-1$, we have 
\be
e_k = \left\{ \begin{array}{ll}
-\frac1{\cZ_{(m;n)}}
\hbar^2 \mu_k^{\partial} \, \cZ_{(m;n)} \,,  &n \leq k \leq 2n-1\\
\\
\left. -\frac1{\cZ_{(m;2n-k)}}
\left(\hbar^2 \mu_k^{\partial} \, 
+\hbar^4 \mu_k^{\partial^2} \, \right) 
 \cZ_{(m;2n-k)}\right|_{\{b_{k>n},c_{k>n}\} \to 0}
\,, & 0 \leq k \leq n-1\\
\\
\left.\nu_k^c -\frac1{\cZ_{(2m+k;n)}}
\left(\hbar^2 \nu_k^{\partial} \, 
+\hbar^4 \nu_k^{\partial^2} \, \right)  \cZ_{(2m+k;n)}
 \right|_{\{b_{-k>m},c_{-k>m}\} \to 0}
 \,, & -(m-1) \leq k \leq -1 \\
\\
\nu_k^c-\frac1{\cZ_{(m;n)}}
\hbar^2 \nu_k^{\partial} \, \cZ_{(m;n)}
 \,, & -(2m-1) \leq k \leq -m  \,.
\end{array} \right.
\label{e_k rep}
\ee
where 
\begin{align}
\begin{split}
\mu_k^\partial=& \sum_{\substack{r+s-t=k \\ t>0}} \frac {t} 2
\left(3  c_r  c_s \frac{\partial}{\partial  c_t}
-6  c_r  b_s \frac{\partial}{\partial  b_t}
-3  b_r  b_s \frac{\partial}{\partial  c_t}
 \right) \\
&-\frac32 \epsilon \sum_{\substack{r-s=k \\s>0}}  \frac{s}2 \left[ (1+r)
\left( \sqrt3  c_r \fpp{ c_s} -2  b_r \fpp{ c_s} 
-\sqrt3  b_r\fpp{ b_s} \right) \right. \\
&\qquad \qquad \qquad \left. +(1-s) 
\left( \sqrt3  c_r \fpp{ c_s} -2  c_r \fpp{ b_s} 
-\sqrt3  b_r\fpp{ b_s} \right) \right]\,,
\end{split} \\
\mu_k^{\partial^2}=&-\sum_{\substack{r-s-t=k\,, \\s,t>0}} \frac{s \, t}4
\left(3  c_r \fpp{ c_s}\fpp{ c_t}-
 6  b_r \fpp{ b_s} \fpp{ c_t}-3  c_r \fpp{ b_s} \fpp{ b_t}
 \right)\,,
\end{align}
\be
\begin{split}
\nu_k^\partial=&
\sum_{ \substack{r+s-t=k \\ t<0}} \frac{(-t)} 2
\left(3  c_r  c_s \frac{\partial}{\partial  c_t}
-6  b_r  c_s \frac{\partial}{\partial  b_t}
-3  b_r  b_s \frac{\partial}{\partial  c_t}
 \right)  \\
&+\epsilon \sum_{\substack{r-s=k \\ s<0}} \frac{(-s)}2
\left(6 c_r N_c \fpp{ c_s}
-6  b_r N_b \fpp{ c_s}-6  b_r N_c \fpp{ b_s}
-6  c_r N_b \fpp{ b_s}
 \right) \\
&-\frac32 \epsilon \sum_{\substack{r-s=k \\ s<0}} \frac{(-s)}2 \left[ (1+r) 
\left( \sqrt3  c_r \fpp{ c_s} -2  b_r \fpp{ c_s} 
-\sqrt3  b_r\fpp{ b_s} \right) \right. \\
&\qquad \qquad  \qquad ~~~~~ \left. + (1-s)
\left( \sqrt3  c_r \fpp{ c_s} -2  c_r \fpp{ b_s}
 -\sqrt3  b_r\fpp{ b_s} \right) \right] \,,
\end{split}
\ee
\begin{align}
\nu_k^{\partial^2}=&-\sum_{\substack{r-s-t=k \\ s,t<0}} \frac{s \, t}4
\left( 3  c_r \fpp{ c_s}\fpp{ c_t}
-6  b_r \fpp{ b_s} \fpp{ c_t}-3  c_r \fpp{ b_s} \fpp{ b_t} \right)\,,
 \\
\begin{split}
\nu_k^c=&+\epsilon \sum_{r+s=k} 
\left( 3 N_c \,  c_r  c_s-6 N_b\,  c_r  b_s  -3 N_c \,  b_r  b_s  \right)
+\epsilon^2 \left(3  c_k N_c^2- 6  b_k N_b N_c-3  c_k N_b^2 \right) 
\\
&-\frac32 \epsilon^2 \left[ (k+1)
\left( \sqrt3  c_k N_c -2  b_k N_c -\sqrt3  b_k N_b \right)+
\left( \sqrt3  c_k N_c-2  c_k N_b -\sqrt3  b_k N_b \right) \right]\,.
\end{split}
\end{align}
Note that the  $\epsilon$ terms  in $\mu_k^\partial$ 
vanished identically when $k \geq n$ 
and same for $\nu_k^\partial$ when  $k \leq -m$.
And $e_{-2m}=\nu_{-2m}^{c}$ is a constant.

Note that we introduced the extended partition function in \eqref{e_k rep};
$\cZ_{(m;2n-k)}$ for $0\leq k \leq n-1$
and $\cZ_{(2m+k;n)}$ for $-(m-1) \leq k \leq -1$.
This is because $g_1$ and $g_2$ can have the expectation 
values $\vev{1/x_i^r}$, $\vev{1/y_j^r}$ 
with $-(2m+k) \leq r < -m$ and $n<r\leq 2n-k$.
To represent these expectation values in terms of derivatives
of the partition function,
we need to extend  the parameter space up to $b_{2n-k}$, $c_{2n-k}$ 
when $0\leq k \leq n-1$,
and up to $b_{-(2m+k)}$ and  $c_{-(2m+k)}$ when $-(m-1)\leq k \leq -1$.
After evaluation of the derivatives, we put the parameters zero \cite{CR_2015}.

\section{Representation of W3 currents}

$\xi_2$ and $\xi_3$ are  the expectation values of
the Virasoro and $\mathcal{W}_3$ current:
\be
\xi_2=\frac{\vev{I_m|T(z)|I_n}}{\vev{I_m|I_n}}
\,, ~~~~~
\xi_3=\frac{\vev{I_m|W(z)|I_n}}{\vev{I_m|I_n}} \,.
\ee
One can check that the modes of $\xi_2$ and $\xi_3$ in \eqref{app_xi}
are compatible with the 
$\mathcal{W}_3$ algebraic commutation relation:

\begin{gather}
[L_p, L_q]=(p-q) L_{p+q}+\frac c{12}(p^3-p) 
\delta_{p,-q} \,, \\
[L_p,W_q]=(2p-q)W_{p+q} \,, 
 \\
\begin{split}
-\frac29\left(\frac{32}{22+5c} \right)[W_p,W_q]
=&\frac{c}{3 \cdot 5 !}(p^2-1)(p^2-4)p \delta_{p,-q}
+\frac{16}{22+5c}(p-q) \Lambda_{p+q} \\
&+(p-q)\left(\frac1{15}(p+q+2)(p+q+3)
-\frac1 6 (p+2)(q+2) \right) L_{p+q}\,,
\end{split} 
\label{app:W3}
\end{gather}
where\footnote{If one rescales
$W_p$ as $i \frac3{\sqrt2}\sqrt{\frac{22+5c}{32}} W_p$, then
the algebra reduces to the original
Fateev and Zamolodchikov convention \cite{FZ_1987}.
Note that because of the factor $i$, the modes of $W(z)$ are anti-hermitian:
$W_k^{\dag}=-W_k$}
\begin{gather}
\Lambda_p=\sum_{k=-\infty}^{\infty} :L_k L_{p-k}:
+\frac15 x_p L_p \,, \nn \\
x_{2\ell}=(\ell+1)(\ell-1) \,,~~~~~ x_{2\ell+1}=(2+\ell)(1-\ell) \nn\,,
\end{gather}
and the central charge $c=2+24 {\epsilon}^2$.

Note that the negative generators $L_{-k}$ and $W_{-k}$ ($k>0$)
obtained in \eqref{app:d_k rep} and \eqref{e_k rep} are left representation  in the sense 
that negative generators should act on the bra  $\langle I_m|$.
However, to check the commutation relation \eqref{app:W3}
we need to find right representations of  negative generators acting 
on  ket $|I_n \rangle$. 
To find the right representation we follow the trick 
used in  \cite{CR_2015}: Use the  transformation of the integration variables 
 $x_i  \to x_i+ \ep/x_i^r$ and $y_j  \to y_j + \ep/y_j^r$
 to obtain two identities  which can be used  to find the relation for $d_k$ ($ k \leq -1$) : 
\be
d_k= 
-\left. \frac{\left( \hbar^2 v_k^\partial +\hbar^4 v_k^{\partial^2} \right) \, \cZ_{(m;n-k)}}
{\cZ_{(m;n-k)}} \right|_{\{b_{k>n},c_{k>n}\} \to 0}    \,,
\ee
where 
\be
v_k^{\partial^2}=-\sum_{-(r+s)=k} \frac{ r\,s}4
\left( \fpp{b_r} \fpp{b_s}+\fpp{c_r} \fpp{c_s} \right)
+\frac{\epsilon}2 k (k+1) \left( \fpp{b_{-k}}+\sqrt3 \fpp{c_{-k}} \right)\,,
\ee
When $k < -m$, one realizes that $d_k$ vanishes identically 
and $v_{-1}^{\partial^2}=0$ by definition.

Likewise, after the change of variables
$x_i \to x_i+\sum_{j=1}^{N_2} \frac{\ep}{(x_i-y_j)x_i^r}$
and
$y_j \to y_j+\sum_{i=1}^{N_1} \frac{\ep}{(x_i-y_j)y_j^r}$,
one finds the right representation of the negative mode $ k \leq -1$  ${\cal W}_3$ current: 
\be
e_k=-\left. \frac{(\hbar^2 \mu_k^\partial +\hbar^4 \, \mu_k^{\partial^2}
+\hbar^6 \, \mu_k^{\partial^3}) \, \cZ_{(m;2n-k)}}{\cZ_{(m;2n-k)}}
\right|_{\{b_{k>n},c_{k>n}\} \to 0} \,,
\ee
where 
\be
\begin{split}
\mu_k^{\partial^3}
=&-\sum_{-(r+s+t)=k}
\frac{r \, s \, t}8 \left(3 \fpp{b_r} \fpp{b_s} \fpp{b_t}
-\fpp{c_r} \fpp{c_s} \fpp{c_t} \right) \\
&+\frac32 \epsilon \sum_{-(r+s)=k} \frac{r \, s \, (1-s)}4
\left(\sqrt3 \fpp{c_r}\fpp{c_s}-2 \fpp{c_r} \fpp{b_s}-\sqrt3 \fpp{b_r}\fpp{b_s} \right)
\\
&+\frac38 \epsilon^2 k(k+1)(k+2)
\left(\sqrt3 \fpp{b_{-k}}-\fpp{c_{-k}} \right)\,.
\end{split}
\ee
We have $e_{k<-2m}=0$ and  $\mu_k^{\partial^3}=0$ for $k=-1,-2$.

If we define the differential operator $v_k$ and $\mu_k$ by
\be
v_k= \left\{ \begin{array}{ll}
v_k^\partial \,,  &-1 \leq k \leq n-1 \\
v_k^\partial+v_k^{\partial^2} \,, &  k \leq -2 
\end{array} \right.
\,, ~~~~~~
\mu_k= \left\{ \begin{array}{ll}
\mu_k^\partial \,,  &n \leq k \leq 2n-1 \\
\mu_k^\partial+\mu_k^{\partial^2} \,, &  -2 \leq k \leq n-1
\\
\mu_k^\partial+\mu_k^{\partial^2}+\mu_k^{\partial^3} \,, &  k \leq -3 \,.
\end{array} \right.
\ee
then, the right representation of the Virasoro and $\mathcal{W}_3$ currents
has the expression 
\be
\mathcal{L}_k = \left\{ \begin{array}{ll}
0 \,,  &2n < k \\
A_k\,, & n \leq k \leq 2n \\
A_k+v_k \,, & -2m \leq k \leq n-1 \\
v_k \,, &  k < -2m  
\end{array} \right.
\,,~~~~
\Omega_k = \left\{ \begin{array}{ll}
0 \,,  &3n < k \\
M_k\,, & 2n \leq k \leq 3n \\
M_k+\mu_k \,, & -3m \leq k \leq 2n-1 \\
\mu_k \,, &  k < -3m  \,.
\end{array} \right.
\ee
where $\vev{I_m|L_k|I_n}:=\mathcal{L}_k \vev{I_m|I_n}$ and 
$\vev{I_m|W_k|I_n}:=\Omega_k \vev{I_m|I_n}$.
One can check that the right representation satisfies the commutation relations
\eqref{app:W3}.

\section{Perturbation method to find flow equations in $A_2$ model}
\label{app:IMM}
In this section, we apply another method to find flow equations, with no need to assume $\epsilon$ to be small. 
Instead, we will suppose hierarchy in the Toda momentum $a_k$ and  $b_k$. 
\subsection{$\cZ_{(0;1)}$}
Expanding in terms of $z$, to the highest power of  \eqref{eq:quadratic},  $ z^{N+M-2}$ shows that
\be 
   d_0 =  {\epsilon}^2\big( \,N(N-1)-NM+M(M-1)\big) +2{\epsilon} [N b_0+Ma_0]\,,
\ee
Expanding \eqref{cubic P} in terms of $z$,  for $ z^{N-3-k}$ we have
 \ba
&& 0= {\epsilon}^3P_{N-k}(N-k)(N-k-1)(N-k-2)\\
&&+2{\epsilon}^2\sum_{ t=-m}^{n} P_{N-k+t}(N-k+t)(N-k-2) (2 b_t +a_t)   
\nn\\
&&+{\epsilon}\bigg( \sum_{ t=-2m}^{2n} P_{N-k+t}(N-k+t)A_t 
-\sum_{ t=-m}^{n-1} P_{N-k+t}(N-k+t)d_t \nn\\
&&+\frac43\sum_{ t=-2m}^{2n} P_{N-k+t}(N-k+t)[\sum_{ s=-m}^{n} (2 b_s +a_s)  (2 b_{t-s} +a_{t-s}) ] \bigg)\nn\\
&&-\sum_{ t=-2m}^{2n-1} P_{N-k+t}[-\frac{2}{3\sqrt3}e_t+\frac23\sum_{ s=-m}^{n-1} d_s (2 b_{t-s} +a_{t-s}) ]+\frac{\epsilon}2 \sum_{ t=-m}^{n-1} P_{N-k+t}(t+2)d_t\,.\nn
\ea

The next power $ z^{N+M-3}$ of   \eqref{eq:quadratic}  gives
 \ba
\label{PQ}
&&  P_{N-1}\bigg\{ {\epsilon}^2\big[M-2(N-1)\big]-2{\epsilon}  b_0\bigg\}+ Q_{M-1}\bigg\{ {\epsilon}^2\big[N-2(M-1)\big]-2{\epsilon}a_0\bigg\}\nn\\
&&=-2{\epsilon}\big[N b_1+M a_1\big]\,,
\ea
Then let's turn back to Eq. \eqref{cubic P}.  For $ z^{N-3-k}$,
\ba
&&0= P_{N-k}\bigg\{- [-\frac{2}{3\sqrt3}e_0+ \frac{2}{3} d_0( 2b_0+a_0)-{\epsilon}d_{0}]+ {\epsilon}^3　(N-k)(N-k-1)(N-k-2)\nn\\
&&+2{\epsilon}^2(N-k)(N-k-2)( 2b_0+a_0)+{\epsilon}(N-k)[-d_0+2{\epsilon}(b_0+a_0)+4( b_0^2+a_0b_0)  ]
\bigg\}\nn\\
&&+P_{N-k+1}\bigg\{- [-\frac{2}{3\sqrt3}e_1+ \frac{2}{3} d_0( 2b_1+a_1)]\\
&&+2{\epsilon}^2(N-k+1)(N-k-2)( 2b_1+a_1)+4{\epsilon}(N-k+1)[{\epsilon}(b_1+a_1)+( 2b_0b_1+a_1b_0+a_0b_1) ]
\bigg\}\nn\\
&&+P_{N-k+2}\bigg\{{\epsilon}(N-k+2)[4( b_1^2+a_1b_1) ]
\bigg\}\,,\nn
\ea
where we have used the definition of $A_k$ in \eqref{akv}.
The corresponding equations of $Q_{M-k}$ can be obtained by setting $P_{N-k} \to Q_{M-k}$, $e_{k} \to -e_{k} $ and $ b_k  \to a_{k}$, $ a_k  \to b_{k}$ .

At each power of z, we have identities,
 \ba
&&  z^{N-3}:  \quad [-\frac{2}{3\sqrt3}e_0+ \frac{2}{3} d_0( 2b_0+a_0)-{\epsilon}d_{0}]= \\
&& \qquad{\epsilon}^3N(N-1)(N-2)
+2{\epsilon}^2N(N-2)( 2b_0+a_0)
++{\epsilon}N[-d_0+2{\epsilon}(b_0+a_0)+4( b_0^2+a_0b_0)]\nn\,,
\ea
%
\ba
\label{zn4}
z^{N-4}:  \quad &&0= P_{N-1}\bigg\{- [-\frac{2}{3\sqrt3}e_0+ \frac{2}{3} d_0( 2b_0+a_0)-{\epsilon}d_{0}]+ {\epsilon}^3　(N-1)(N-2)(N-3)\nn\\
&&+2{\epsilon}^2(N-1)(N-3)( 2b_0+a_0)+{\epsilon}(N-1)[-d_0+2{\epsilon}(b_0+a_0)+4( b_0^2+a_0b_0)  ]
\bigg\}\nn\\
&&+\bigg\{- [-\frac{2}{3\sqrt3}e_1+ \frac{2}{3} d_0( 2b_1+a_1)]\\
&&+2{\epsilon}^2N(N-3)( 2b_1+a_1)+4{\epsilon}N[{\epsilon}(b_1+a_1)+( 2b_0b_1+a_1b_0+a_0b_1) ]
\bigg\}\nn\,,
\ea
\ba
\label{zn5}
z^{N-5}:  \quad&&0= P_{N-2}\bigg\{- [-\frac{2}{3\sqrt3}e_0+ \frac{2}{3} d_0( 2b_0+a_0)-{\epsilon}d_{0}]+ {\epsilon}^3　(N-2)(N-3)(N-4)\nn\\
&&+2{\epsilon}^2(N-2)(N-4)( 2b_0+a_0)+{\epsilon}(N-2)[-d_0+2{\epsilon}(b_0+a_0)+4( b_0^2+a_0b_0)  ]
\bigg\}\nn\\
&&+P_{N-1}\bigg\{- [-\frac{2}{3\sqrt3}e_1+ \frac{2}{3} d_0( 2b_1+a_1)]\\
&&+2{\epsilon}^2(N-1)(N-4)( 2b_1+a_1)+4{\epsilon}(N-1)[{\epsilon}(b_1+a_1)+( 2b_0b_1+a_1b_0+a_0b_1) ]
\bigg\}\nn\\
&&+\bigg\{{\epsilon}N[4( b_1^2+a_1b_1) ]
\bigg\}\,,\nn
\ea
To find $e_1$, we use perturbation assuming  $\vert{ b_1} \vert  \ll \vert{ a_1} \vert$ so that $\vert P_{N-k}\vert \sim \vert{ b_1}\vert^k$.
Then at the first order  we have 
from the above equations \eqref{zn4} and \eqref{zn5}:
\ba
 z^{N-4}:  \quad  \frac{2}{3\sqrt3}e_1^{(1)}
&&=\frac{2}{3} d_0a_1-2{\epsilon}^2N(N-1) a_1-4{\epsilon}N b_0a_1\equiv B_1( b_0, a_0) a_1
\,,
\ea
\ba
z^{N-5}:  \quad&&0=P_{N-1}^{(1)}\bigg\{- [-\frac{2}{3\sqrt3}e_1^{(1)}+ \frac{2}{3} d_0a_1]\\
&&+2{\epsilon}^2(N-1)(N-4)a_1+4{\epsilon}(N-1)[{\epsilon}a_1+a_1b_0]
\bigg\}+\bigg\{{\epsilon}N[4a_1b_1 ]
\bigg\}\,,\nn
\ea
\ba
P_{N-1}^{(1)}=\frac{Nb_1}{\epsilon(N-1)+b_0}\,.
\ea
At the second order, we have
\ba
z^{N-4}:  \quad &&0= P_{N-1}^{(1)}\bigg\{- [-\frac{2}{3\sqrt3}e_0+ \frac{2}{3} d_0( 2b_0+a_0)-{\epsilon}d_{0}]+ {\epsilon}^3　(N-1)(N-2)(N-3)\nn\\
&&+2{\epsilon}^2(N-1)(N-3)( 2b_0+a_0)+{\epsilon}(N-1)[-d_0+2{\epsilon}(b_0+a_0)+4( b_0^2+a_0b_0)  ]
\bigg\}\nn\\
&&+\bigg\{- [-\frac{2}{3\sqrt3}e_1^{(2)} + \frac{4}{3} d_0b_1]+4{\epsilon}^2N(N-3)b_1+4{\epsilon}N[{\epsilon}b_1+( 2b_0b_1+a_0b_1) ]
\bigg\}\,,
\ea
\ba
 &&\frac{2}{3\sqrt3}e_1^{(2)} =\frac{4}{3} d_0b_1 -4{\epsilon}N( 2b_0+a_0)b_1-4{\epsilon}^2N(N-2)b_1 +\frac{Nb_1}{\epsilon(N-1)+b_0}\bigg\{4{\epsilon}( b_0+a_0) b_0\nn\\
&&\qquad \quad \quad+2{\epsilon}^2( b_0+a_0) +2{\epsilon}^2(2N-3)( 2b_0+a_0) +3{\epsilon}^3　(N-1)(N-2)-{\epsilon}d_0
\bigg\}\\
&&\qquad \quad \quad \equiv B_2( b_0, a_0) b_1\nn \,.
\ea
Up to $\cO(b_1)$ we find $e_1=e_1^{(1)}+e_1^{(2)}$. If one expands $e_1$ in terms of $\epsilon$, it reads
 \ba
&&\frac{2}{3\sqrt3}e_1=\frac{4}{3} \epsilon \left[ (a_1+2 b_1)(a_0 M+ b_0 N)
-3 b_0 N(a_1+b_1)  \right]
\\
&&+  \epsilon^2
\left[\frac{2}{3}(a_1+2 b_1) (N(N-1) + M(M-1) -NM)-2N(N-1)(a_1+b_1)-2NM\frac{a_0}{b_0} b_1 \right]
+\cO(\epsilon^3).\nn
\ea
This result is in perfect agreement with \eqref{e1e}, expanding up to $\cO(b_1)$. In fact, the perturbative condition $\vert{ b_1} \vert  \ll \vert{ a_1} \vert$ is equivalent to choosing the filling fraction $N_1=N$, and $N_2=0$.

In this way we can find that 
 \ba
\label{ev}
&&e_1=e_1^{(1)}+e_1^{(2)} +e_1^{(3)} +...+e_1^{(k+2)} +...  \\
&&=B_1( b_0, a_0) a_1+B_2( b_0, a_0) b_1+B_3( b_0, a_0) b_1\frac{b_1}{a_1}+...+B_{k+2}( b_0, a_0) b_1(\frac{b_1}{ a_1})^k +...\nn
\ea
The flow equations for rank 1 case are
\be
\label{doe1}
-\hbar^2v_0 \log \cZ_{1}=d_0 \,,~~~~
-\hbar^2\mu_1 \log \cZ_{1}=e_1\,,
\ee
where $v_0= b_1 \fpp{ b_1}+a_1 \fpp{a_1}$ and
$\mu_1 =  
{\sqrt3}( a_1^2 +2a_1  b_1 )  \frac{\partial}{\partial a_1}
- \sqrt3( 2 a_1 b_1 +b_1^2) \frac{\partial}{\partial b_1} 
$. From the first equation of \eqref{doe1} we find
\be
\hbar^2\log \cZ_{1}=-d_0 \log a_1+H(t)\,,
\ee
where $t:= b_1/a_1$ and  $H(t)$ is a homogeneous solution to $v_0$. Put $H(t)$ into the second equation
of \eqref{doe1},  we get
\be
3(t+1)t \frac{\partial H(t)}{\partial t}
=\frac1{\sqrt3}\frac{e_1}{a_1}-(1+2t)d_0 \,.
\ee

From \eqref{ev}, it is clear that 
 \ba
\frac{e_1}{a_1}=B_1( b_0, a_0) +B_2( b_0, a_0) t+B_3( b_0, a_0) t^2+...
\ea
Therefore, we have
\begin{align}
 H(t)
&=\frac1{3}(\frac {B_1}{\sqrt3} -d_0)\log t-\frac1{3}(\frac {B_1-B_2+B_3}{\sqrt3} +d_0)\log (t+1)
+\frac {B_3}{3\sqrt3} t+...\,,
\end{align}
and partition function
\begin{align}
\cZ_{(0;1)} & =\mathcal{N} a_1^{-{d_0}/{\hbar^2}}t^{(\frac {B_1}{\sqrt3} -d_0)/3\hbar^2}
{(t-1)}^{-(\frac {B_1-B_2+B_3}{\sqrt3} +d_0)/3\hbar^2} e^{{B_3t}/{(3\sqrt3)} }\,,
\end{align}
where $\mathcal{N}$ is a function of $a_0$, $b_0$  and $ B_k$ with $k \geq 4$.

\subsection{$\cZ_{(0;2)}$}
We need $d_0$, $d_1$, $e_2$ and  $e_3$ to obtain the partition function.
From the highest power $ z^{N+M-2}$ of the quadratic equation\eqref{+ek}  we obtain the expression of $d_0$ for any rank n.
Now for rank 2 case, from the second highest power $ z^{N+M-3}$ we have
 \ba
&&  P_{N-1}\bigg\{ -d_0
+{\epsilon}^2\big[(N-1)(N-2)-(N-1)M+(M-1)M\big]
+2{\epsilon}\big[(N-1) b_0+M  a_0\big]\bigg\}\nn\\
&& + Q_{M-1}\bigg\{ -d_0
+{\epsilon}^2\big[(M-1)(M-2)-(M-1)N+(N-1)N\big]
+2{\epsilon}\big[N b_0+(M-1) a_0\big]\bigg\}\nn\\
&&=d_1-2{\epsilon}\big[N  b_1+M a_1\big]\,,
\ea

From the cubic equation \eqref {cubic P}, we have
 \ba
&&z^{N-3}:  [-\frac{2}{3\sqrt3}e_0+\frac23d_0 (2 b_{0} +a_{0})-{\epsilon}d_0 ]
\\
&&= {\epsilon}^3N(N-1)(N-2)+2{\epsilon}^2 N(N-2) (2 b_0 +a_0)
+{\epsilon} N(A_0-d_0) +\frac43\epsilon N (2 b_0 +a_0)^2  \nn\,,
\ea
 \ba
\label{znn4}  
&&z^{N-4}:
P_{N-1}[-\frac{2}{3\sqrt3}e_0+\frac23d_0 (2 b_{0} +a_{0})-{\epsilon}d_0 ]
\nn\\
&&+ [-\frac{2}{3\sqrt3}e_1+\frac23d_0 (2 b_{1} +a_{1})+\frac23d_1 (2 b_{0} +a_{0})-\frac32{\epsilon}d_1 ]\nn\\
&&= {\epsilon}^3P_{N-1}(N-1)(N-2)(N-3)\\
&&+2{\epsilon}^2\bigg( P_{N-1}(N-1)(N-3) (2 b_0 +a_0)  + N(N-3) (2 b_1 +a_1)\bigg) 
\nn\\
&&+{\epsilon}\bigg(  P_{N-1}(N-1)(A_0-d_0)+N(A_1-d_1)\bigg) \nn\\
&&+\frac43\epsilon\bigg(P_{N-1}(N-1) (2 b_0 +a_0)^2  +
2N(2 b_0 +a_0) (2 b_1 +a_{1}) \bigg)\nn\,,
\ea
 \ba
&&z^{N-5}:  P_{N-2}[-\frac{2}{3\sqrt3}e_0+\frac23d_0 (2 b_{0} +a_{0})-{\epsilon}d_0 ]
\nn\\
&&+ P_{N-1}[-\frac{2}{3\sqrt3}e_1+\frac23d_0 (2 b_{1} +a_{1})+\frac23d_1 (2 b_{0} +a_{0})-\frac32{\epsilon}d_1 ]\nn\\
&&+ [-\frac{2}{3\sqrt3}e_2+\frac23d_0 (2 b_{2} +a_{2})+\frac23d_1 (2 b_{1} +a_{1})]\nn\\
&&= {\epsilon}^3P_{N-2}(N-2)(N-3)(N-4)\\
&&+2{\epsilon}^2\bigg( P_{N-2}(N-2)(N-4) (2 b_0 +a_0)  + P_{N-1}(N-1)(N-4) (2 b_1 +a_1)\nn\\
&&+ N(N-4) (2 b_2 +a_2)\bigg) 
\nn\\
&&+{\epsilon}\bigg(  P_{N-2}(N-2)(A_0-d_0)+P_{N-1}(N-1)(A_1-d_1)+NA_2\bigg) \nn\\
&&+\frac43\epsilon\bigg(P_{N-2}(N-2) (2 b_0 +a_0)^2  +
2P_{N-1}(N-1)(2 b_0 +a_0) (2 b_1 +a_{1}) \nn\\
&&+ N[ 2(2 b_0 +a_0)(2 b_{2} +a_2)+(2 b_1 +a_1)^2  ]  \bigg)\nn\,,
\ea
 \ba
&&z^{N-6}:  P_{N-3}[-\frac{2}{3\sqrt3}e_0+\frac23d_0 (2 b_{0} +a_{0})-{\epsilon}d_0 ]
\nn\\
&&+ P_{N-2}[-\frac{2}{3\sqrt3}e_1+\frac23d_0 (2 b_{1} +a_{1})+\frac23d_1 (2 b_{0} +a_{0})-\frac32{\epsilon}d_1 ]\nn\\
&&+ P_{N-1}[-\frac{2}{3\sqrt3}e_2+\frac23d_0 (2 b_{2} +a_{2})+\frac23d_1 (2 b_{1} +a_{1})]\nn\\
&&+[-\frac{2}{3\sqrt3}e_3+\frac23d_1 (2 b_{2} +a_{2})]\nn\\
&&= {\epsilon}^3P_{N-3}(N-3)(N-4)(N-5)\\
&&+2{\epsilon}^2\bigg( P_{N-3}(N-3)(N-5) (2 b_0 +a_0)  + P_{N-2}(N-2)(N-5) (2 b_1 +a_1)\nn\\
&&+ P_{N-1}(N-1)(N-5) (2 b_2 +a_2)\bigg) 
\nn\\
&&+{\epsilon}\bigg(  P_{N-3}(N-3)(A_0-d_0)+P_{N-2}(N-2)(A_1-d_1)+P_{N-1}(N-1)A_2\nn\\
&&+4N(2b_1b_2+a_1b_2+a_2b_1)\bigg) \nn\\
&&+\frac43\epsilon\bigg(P_{N-3}(N-3) (2 b_0 +a_0)^2  +
2P_{N-2}(N-2)(2 b_0 +a_0) (2 b_1 +a_{1}) \nn\\
&&+ P_{N-1}(N-1)[ 2(2 b_0 +a_0)(2 b_{2} +a_2)+(2 b_1 +a_1)^2  ]  \bigg)\nn\,,
\ea
 \ba
\label{znn7} 
&&z^{N-7}:  P_{N-4}[-\frac{2}{3\sqrt3}e_0+\frac23d_0 (2 b_{0} +a_{0})-{\epsilon}d_0 ]
\nn\\
&&+ P_{N-3}[-\frac{2}{3\sqrt3}e_1+\frac23d_0 (2 b_{1} +a_{1})+\frac23d_1 (2 b_{0} +a_{0})-\frac32{\epsilon}d_1 ]\nn\\
&&+ P_{N-2}[-\frac{2}{3\sqrt3}e_2+\frac23d_0 (2 b_{2} +a_{2})+\frac23d_1 (2 b_{1} +a_{1})]\nn\\
&&+ P_{N-1}[-\frac{2}{3\sqrt3}e_3+\frac23d_1 (2 b_{2} +a_{2})]\nn\\
&&= {\epsilon}^3P_{N-4}(N-4)(N-5)(N-6)\\
&&+2{\epsilon}^2\bigg( P_{N-4}(N-4)(N-6) (2 b_0 +a_0)  + P_{N-3}(N-3)(N-6) (2 b_1 +a_1)\nn\\
&&+ P_{N-2}(N-2)(N-6) (2 b_2 +a_2)\bigg) 
\nn\\
&&+{\epsilon}\bigg(  P_{N-4}(N-4)(A_0-d_0)+P_{N-3}(N-3)(A_1-d_1)+P_{N-2}(N-2)A_2\nn\\
&&+P_{N-1}(N-1)A_3+4Nb_2( b_2 +a_2)\bigg) \nn\\
&&+\frac43\epsilon\bigg(P_{N-4}(N-4) (2 b_0 +a_0)^2  +
2P_{N-3}(N-3)(2 b_0 +a_0) (2 b_1 +a_{1}) \nn\\
&&+ P_{N-2}(N-2)[ 2(2 b_0 +a_0)(2 b_{2} +a_2)+(2 b_1 +a_1)^2  ] \nn\\
&&+2P_{N-1}(N-1)(2 b_1 +a_1) (2 b_2 +a_2) \bigg)\nn\,.
\ea
Perturbation holds if  we  require $\vert{ b_2}/{b_1} \vert \ll \vert{ b_1} \vert  \ll1$, $\vert{ b_2} \vert  \ll \vert{ a_2} \vert$,  and $\vert{ b_1}\vert \sim  \vert{ a_1} \vert$, so that $\vert P_{N-k}\vert \sim \vert{ b_2}/{b_1} \vert^k$ is ensured.
Then at the first order of the perturbation, we have:\\
From the quadratic equation:
\be 
   d_0 =  {\epsilon}^2\big( \,N(N-1)-NM+M(M-1)\big) +2{\epsilon} [N b_0+Ma_0]\,,
\ee
 \ba
z^{N+M-3}:  \qquad d_1^{(1)}=2{\epsilon}\big[N b_1+M a_1\big]\,.
\ea
From the  equations \eqref{znn4}  to \eqref{znn7}:
 \ba
&&z^{N-4}:  [-\frac{2}{3\sqrt3}e_1^{(1)}+\frac23d_0 (2 b_{1} +a_{1})+\frac23d_1^{(1)} (2 b_{0} +a_{0})-\frac32{\epsilon}d_1^{(1)} ]\nn\\
&&=2{\epsilon}^2N(N-3) (2 b_1 +a_1)
+{\epsilon}N(A_1-d_1^{(1)}) +\frac83\epsilon
N(2 b_0 +a_0) (2 b_1 +a_{1}) \,,
\ea
 \ba
&&z^{N-5}:  [-\frac{2}{3\sqrt3}e_2^{(1)}+\frac23d_0 a_{2}+\frac23d_1^{(1)} (2 b_{1} +a_{1})]\\
&&= 2{\epsilon}^2N(N-4)a_2
+{\epsilon}NA_2+\frac43\epsilon N[ 2(2 b_0 +a_0)a_2+(2 b_1 +a_1)^2  ]  \nn\,,
\ea
 \ba
&&z^{N-6}:  [-\frac{2}{3\sqrt3}e_3^{(1)}+\frac23d_1^{(1)}a_{2}  ] =4{\epsilon}Na_2b_1\,,
\ea

 \ba
&&z^{N-7}:  P_{N-1}^{(1)}[-\frac{2}{3\sqrt3}e_3^{(1)}+\frac23d_1^{(1)} a_{2}]\\
&&= {\epsilon}\bigg( P_{N-1}^{(1)}(N-1)A_3+4Nb_2a_2\bigg) +\frac83\epsilon P_{N-1}(N-1)(2 b_1 +a_1)a_2\nn\,.
\ea
Thus $d_0$, $d_1$, $e_2$ and  $e_3$ are obtained.
\subsection{$\cZ_{(1;1)}$}
We need $d_0$, $e_{-1}$ and  $e_1$ to evaluate.
From the power expansion of the quadratic equation, we know
 \be
z^{N+M-1}:  \qquad d_{-1}=2{\epsilon}\big[N b_{-1}+M a_{-1}\big]\,,
\ee
\ba
&&z^{N+M-2}:  \qquad d_{-1}(P_{N-1}+Q_{M-1})+d_0\\
&&\qquad ={\epsilon}^2\big( \,N(N-1)-NM+M(M-1)\big) +2{\epsilon} [N b_0+Ma_0]\nn\\
&&\qquad+2{\epsilon} [  b_{-1}\big((N-1)P_{N-1}+NQ_{M-1}\big)
+a_{-1}\big((M-1)Q_{M-1}+MP_{N-1}\big)]\nn\,,
\ea
\ba
&&z^{N+M-3}:  \qquad d_{-1}(P_{N-2}+Q_{M-2}+P_{N-1}Q_{M-1})+d_0(P_{N-1}+Q_{M-1})\\
&&= P_{N-1}\bigg\{ 
{\epsilon}^2\big[(N-1)(N-2)-(N-1)M+(M-1)M\big]
+2{\epsilon}\big[(N-1)  b_0+M a_0\big]\bigg\}\nn\\
&& + Q_{M-1}\bigg\{ 
{\epsilon}^2\big[(M-1)(M-2)-(M-1)N+(N-1)N\big]
+2{\epsilon}\big[N  b_0+(M-1) a_0\big]\bigg\}\nn\\
&&+2{\epsilon} [  b_{-1}\big((N-2)P_{N-2}+NQ_{M-2}+(N-1)P_{N-1}Q_{M-1}\big)\nn\\
&&+a_{-1}\big((M-2)Q_{M-2}+MP_{N-2}+(M-1)Q_{M-1}P_{N-1}\big)]\nn\,,
\ea
From the cubic equation
\ba
&&  z^{N-1}:  \quad    [ -\frac{2}{3\sqrt3} e_{-2}+ \frac{2}{3} d_{-1}(2b_{-1}+a_{-1})] 
= \frac43{\epsilon}N(2b_{-1}+a_{-1})^2
+{\epsilon}NA_{-2}
\,,
\ea
\ba
&&  z^{N-2}:  \quad    P_{N-1} [ -\frac{2}{3\sqrt3} e_{-2}+ \frac{2}{3} d_{-1}(2b_{-1}+a_{-1})] \nn\\
&&+ [-\frac{2}{3\sqrt3}e_{-1}+ \frac{2}{3} d_{0}(2b_{-1}+a_{-1})+ \frac{2}{3} d_{-1}(2b_{0}+a_{0})-\frac{\epsilon}{2}d_{-1}]\nn\\
&&
=  2{\epsilon}^2N(N-1)(2b_{-1}+a_{-1})
\nn\\
&&+\frac43{\epsilon}\bigg(
2N(2b_{-1}+a_{-1})(2b_{0}+a_{0})+P_{N-1}(N-1)(2b_{-1}+a_{-1})^2\bigg)\nn\\
&&
+{\epsilon}\bigg(P_{N-1}(N-1)A_{-2}+ N(A_{-1}-d_{-1})
\bigg)\,,\nn
\ea
{\small
\ba
&&  z^{N-3}:  \quad    P_{N-2} [ -\frac{2}{3\sqrt3} e_{-2}+ \frac{2}{3} d_{-1}(2b_{-1}+a_{-1})] \nn\\
&&+ P_{N-1}[-\frac{2}{3\sqrt3}e_{-1}+ \frac{2}{3} d_{0}(2b_{-1}+a_{-1})+ \frac{2}{3} d_{-1}(2b_{0}+a_{0})-\frac{\epsilon}{2}d_{-1}]\nn\\
&&
+  [-\frac{2}{3\sqrt3}e_0+ \frac{2}{3} d_{0}(2b_{0}+a_{0})+\frac{2}{3} d_{-1}(2b_{1}+a_{1})-{\epsilon}d_{0}]
\nn\\
&&=   {\epsilon}^3N(N-1)(N-2)\\
&&+2{\epsilon}^2\bigg(N(N-2)(2b_{0}+a_{0}) +P_{N-1}(N-1)(N-2)(2b_{-1}+a_{-1})
\bigg)\nn\\
&&+\frac43{\epsilon}\bigg(
N(2b_{0}+a_{0})^2+2P_{N-1}(N-1)(2b_{-1}+a_{-1})(2b_{0}+a_{0})+P_{N-2}(N-2)(2b_{-1}+a_{-1})^2\bigg)\nn\\
&&
+{\epsilon}\bigg(P_{N-2}(N-2)A_{-2}+ P_{N-1}(N-1)(A_{-1}-d_{-1})+N(A_0-d_0)
\bigg)\,,\nn
\ea
\ba
&&  z^{N-4}:  \quad    P_{N-3} [ -\frac{2}{3\sqrt3} e_{-2}+ \frac{2}{3} d_{-1}(2b_{-1}+a_{-1})] \nn\\
&&+ P_{N-2}[-\frac{2}{3\sqrt3}e_{-1}+ \frac{2}{3} d_{0}(2b_{-1}+a_{-1})+ \frac{2}{3} d_{-1}(2b_{0}+a_{0})-\frac{\epsilon}{2}d_{-1}]\nn\\
&&
+  P_{N-1}[-\frac{2}{3\sqrt3}e_0+ \frac{2}{3} d_{0}(2b_{0}+a_{0})+\frac{2}{3} d_{-1}(2b_{1}+a_{1})-{\epsilon}d_{0}]
+ [-\frac{2}{3\sqrt3}e_1+ \frac{2}{3} d_{0}(2b_{1}+a_{1})]\nn\\
&&=  {\epsilon}^3P_{N-1}(N-1)(N-2)(N-3)\\
&&+2{\epsilon}^2\bigg(N(N-3)(2b_{1}+a_{1})+P_{N-1}(N-1)(N-3)(2b_{0}+a_{0}) +P_{N-2}(N-2)(N-3)(2b_{-1}+a_{-1})
\bigg)\nn\\
&&+\frac43{\epsilon}\bigg( 2N(2b_{0}+a_{0})(2b_{1}+a_{1})
+P_{N-1}(N-1)(2b_{0}+a_{0})^2+2P_{N-2}(N-2)(2b_{-1}+a_{-1})(2b_{0}+a_{0})\nn\\
&&+P_{N-3}(N-3)(2b_{-1}+a_{-1})^2\bigg)\nn\\
&&
+{\epsilon}\bigg(P_{N-3}(N-3)A_{-2}+ P_{N-2}(N-2)(A_{-1}-d_{-1})+P_{N-1}(N-1)(A_0-d_0)
+NA_1\bigg)\,,\nn
\ea

\ba
&&  z^{N-5}:  \quad    P_{N-4} [ -\frac{2}{3\sqrt3} e_{-2}+ \frac{2}{3} d_{-1}(2b_{-1}+a_{-1})] \nn\\
&&+ P_{N-3}[-\frac{2}{3\sqrt3}e_{-1}+ \frac{2}{3} d_{0}(2b_{-1}+a_{-1})+ \frac{2}{3} d_{-1}(2b_{0}+a_{0})-\frac{\epsilon}{2}d_{-1}]\nn\\
&&
+  P_{N-2}[-\frac{2}{3\sqrt3}e_0+ \frac{2}{3} d_{0}(2b_{0}+a_{0})+\frac{2}{3} d_{-1}(2b_{1}+a_{1})-{\epsilon}d_{0}]
+P_{N-1} [-\frac{2}{3\sqrt3}e_1+ \frac{2}{3} d_{0}(2b_{1}+a_{1})]\nn\\
&&=  {\epsilon}^3P_{N-2}(N-2)(N-3)(N-4)\\
&&+2{\epsilon}^2\bigg(P_{N-1}(N-1)(N-4)(2b_{1}+a_{1})+P_{N-2}(N-2)(N-4)(2b_{0}+a_{0}) \nn\\
&&+P_{N-3}(N-3)(N-4)(2b_{-1}+a_{-1})
\bigg)\nn\\
&&+\frac43{\epsilon}\bigg( 2P_{N-1}(N-1)(2b_{0}+a_{0})(2b_{1}+a_{1})
+P_{N-2}(N-2)(2b_{0}+a_{0})^2\nn\\
&&+2P_{N-3}(N-3)(2b_{-1}+a_{-1})(2b_{0}+a_{0})+P_{N-4}(N-4)(2b_{-1}+a_{-1})^2\bigg)\nn\\
&&
+{\epsilon}\bigg(P_{N-4}(N-4)A_{-2}+ P_{N-3}(N-3)(A_{-1}-d_{-1})+P_{N-2}(N-2)(A_0-d_0)
+P_{N-1}(N-1)A_1\bigg)\nn\\
&&+4{\epsilon}N(b_{1}^2+a_{1}b_{1})\,,\nn
\ea}
The corresponding equations of $Q_{M-k}$ can be obtained by setting $P_{N-k} \to Q_{M-k}$ , $e_{k} \to -e_{k} $ and $ b_k  \to a_{k}$.
Again we can apply perturbation method.\\
To apply perturbation, we assume  $\vert{ b_1} \vert  \ll \vert{ a_1} \vert$,  $\vert{ b_1}{ b_{-1}} \vert  \ll1$ and $\vert{a_1}{a_{-1}} \vert  \ll1$ so that $\vert P_{N-k}\vert \sim \vert{ b_1}\vert^k$ is ensured.
\footnote{ Although under this condition $ Q_{M-k}$ cannot be found by perturbation, $d_k$ and $e_k$ can be totally fixed by symmetry.  Notice that $d_k$ is invariant under the transformation $N \to M$  and $ a_{l}  \to b_l$, while $e_k$ is anti-invariant when $N \to M$  and $ a_{l}  \to b_l$. Thus  the explicit dependence of $P_{N-k}$ is enough to determine $d_k$ and $e_k$.} 
Then at the first order, we have\\
from the quadratic equation:
\be 
z^{N+M-2}: \qquad    d_0^{(1)} =  {\epsilon}^2\big( \,N(N-1)-NM+M(M-1)\big) +2{\epsilon} [N b_0+M a_0]\,,
\ee
From the cubic equation:
\ba
&&  z^{N-2}:  \quad[-\frac{2}{3\sqrt3}e_{-1}^{(1)}+ \frac{2}{3} d_{0}^{(1)}(2b_{-1}+a_{-1})+ \frac{2}{3} d_{-1}(2b_{0}+a_{0})-\frac{\epsilon}{2}d_{-1}]\\
&&
=  2{\epsilon}^2N(N-1)(2b_{-1}+a_{-1})
+\frac83{\epsilon}
N(2b_{-1}+a_{-1})(2b_{0}+a_{0})
+{\epsilon} N(A_{-1}-d_{-1})\,,\nn
\ea
\be
\begin{split}
e_{-1}^{(1)}= &
4\sqrt3\epsilon 
\left[ M( \, a_0 a_{-1}+a_0 b_{-1}+b_0 a_{-1})- N( \, b_0 b_{-1}+a_0 b_{-1}+b_0 a_{-1} )  \right] \\
+&\sqrt3\epsilon^2 \left[a_{-1}( M^2-2N^2+2NM+2N-\frac52 M)+
b_{-1}( 2M^2-N^2-2NM-2M+\frac52 N)
 \right]\,,
\end{split}
\ee
\ba
&&  z^{N-3}:  \quad    [-\frac{2}{3\sqrt3}e_0^{(1)}+ \frac{2}{3} d_{0}^{(1)}(2b_{0}+a_{0})-{\epsilon}d_{0}^{(1)}]
\\
&&=   {\epsilon}^3N(N-1)(N-2)+2{\epsilon}^2N(N-2)(2b_{0}+a_{0}) +\frac43{\epsilon}
N(2b_{0}+a_{0})^2
+{\epsilon}N(A_0-d_0)
\,,\nn
\ea
\ba
  z^{N-4}:  \quad     [-\frac{2}{3\sqrt3}e_1^{(1)}+ \frac{2}{3} d_{0}^{(1)}a_{1}]=  2{\epsilon}^2N(N-3)a_{1}+4{\epsilon}^2Na_{1}+4{\epsilon}Nb_{0}a_{1}\,,
\ea
\ba
e_1^{(1)}= a_{1}\bigg( 2\sqrt3{\epsilon}[Ma_{0}-2Nb_{0}] +\sqrt3{\epsilon}^2[-2N(N-1)-NM+M(M-1)]\bigg)\,,
\ea

\ba
&&  z^{N-5}:  \quad   
P_{N-1}^{(1)} [-\frac{2}{3\sqrt3}e_1^{(1)}+ \frac{2}{3} d_{0}^{(1)}a_{1}]-4{\epsilon}Na_{1}b_{1}\\
&&= P_{N-1}^{(1)}\bigg(2{\epsilon}^2(N-1)(N-4)a_{1}+\frac83{\epsilon}(N-1)(2b_{0}+a_{0})a_{1}
+{\epsilon}(N-1)(4{\epsilon}a_1-\frac43(2a_{0}a_{1}+b_{0}a_{1}))\bigg)\,,\nn
\ea
\ba
P_{N-1}^{(1)}=\frac{Nb_1}{3\epsilon(N-1)+b_0}\,.
\ea
Second order contribution is given as follows:
\ba
z^{N+M-2}:  \qquad d_0^{(2)} =-2{\epsilon} [  b_{-1}P_{N-1}^{(1)}
+a_{-1}Q_{M-1}^{(1)}]\,,
\ea

\be
d_0^{(2)} =-2{\epsilon}\left(
\frac{ Nb_1 b_{-1}}{3\epsilon(N-1)+b_0}  + \frac{Ma_1 a_{-1}}{3\epsilon(M-1)+a_0} \right)\,.
\ee

\ba
  z^{N-2}:  \quad    P_{N-1} [\frac43{\epsilon}(2b_{-1}+a_{-1})^2+{\epsilon}NA_{-2}]+ [-\frac{2}{3\sqrt3}e_{-1}^{(2)}+ \frac{2}{3} d_{0}^{(2)}(2b_{-1}+a_{-1})]=0\,,
\ea

\be
e_{-1}^{(2)}=2\sqrt3 {\epsilon}\left(
\frac{(2 a_{-1}+b_{-1}) N}{3\epsilon(N-1)+b_0} b_1 b_{-1} - \frac{(a_{-1}+2b_{-1})M}{3\epsilon(M-1)+a_0} a_1 a_{-1}\right)\,.
\ee
\ba
&&  z^{N-4}:  \quad     P_{N-1}^{(1)} [-\frac{2}{3\sqrt3}e_0^{(1)}+ \frac{2}{3} d_{0}^{(1)}(2b_{0}+a_{0})-{\epsilon}d_{0}^{(1)}]
+ [-\frac{2}{3\sqrt3}e_1^{(2)}+ \frac{4}{3} d_{0}^{(1)}b_{1}]\nn\\
&&=  {\epsilon}^3P_{N-1}^{(1)}(N-1)(N-2)(N-3)\\
&&+2{\epsilon}^2\bigg(N(N-3)2b_{1}+P_{N-1}^{(1)}(N-1)(N-3)(2b_{0}+a_{0}) 
\bigg)\nn\\
&&+\frac43{\epsilon}\bigg( 4N(2b_{0}+a_{0})b_{1}
+P_{N-1}^{(1)}(N-1)(2b_{0}+a_{0})^2\bigg)\nn\\
&&
+{\epsilon}\bigg(P_{N-1}^{(1)}(N-1)(A_0-d_0)
+N[4{\epsilon}b_{1}-\frac43(2b_{0}+a_{0})b_{1}]\bigg)\,,\nn
\ea

\ba
&&e_1^{(2)}=2\sqrt3 b_{1}\bigg( {\epsilon}[2Ma_{0}-3Na_{0}-4Nb_{0}] +{\epsilon}^2[-2N^2+5N-NM+M(M-1)]\bigg)\nn\\
&&+\frac{3\sqrt3}{2}\frac{Nb_1}{3\epsilon(N-1)+b_0}\bigg( 4{\epsilon}(a_{0}+b_{0})b_{0}+2{\epsilon}^2[b_{0}(3N-5)+a_{0}(2N-2-M)] \nn\\
&&+{\epsilon}^3 [2(N-1)(N-3)-NM+M(M-1)]\bigg)
\,.
\ea
From this consideration we have
\ba
d_0=d_0^{(1)}+d_0^{(2)} +... =  D_0 +D_1b_{-1} b_{1}+D_2a_{-1} a_{1}+ higher \,order \,,
\ea
\ba
e_1=e_1^{(1)}+e_1^{(2)} +... =D_3a_{1}+D_4b_{1}+ higher \,order\,,
\ea
\ba
&& e_{-1} =e_{-1} ^{(1)}+e_{-1} ^{(2)} +...\\
 &&=  D_5  b_{-1}+ D_6 a_{-1}+D_7{(2 a_{-1}+b_{-1}) }b_1 b_{-1} + D_8{(a_{-1}+2b_{-1})} a_1 a_{-1}+ higher \,order\nn\,,
\ea
where all $D_k$ are functions of $b_0$, $a_0$, which can be read off from the above equations.
We compare these results with the $\epsilon$ expansion \eqref{d011}-\eqref{e-111} and find
 they agree with each other,
by calculating  \eqref{d011}-\eqref{e-111} further to order 
 $\cO(\eta_0^2)$ with $M=M_1$ and $N=N_1$:
\be
d_0^{(1;1)}=2 \epsilon(b_0  N+a_0 M)-2  \epsilon
 \left(\frac{M}{a_0} a_1 a_{-1}+\frac{N}{b_0} b_1 b_{-1} \right)
 +\cO(\eta_0^2)\,,
\ee
\be
\begin{split}
e_1^{(1;1)}=&
2\sqrt3 \epsilon \left[a_0 (a_1+2 b_1) M-b_0 ( 2 a_1+b_1)N \right]
\\
&-2\sqrt3 \epsilon \left( \frac{(a_1+2b_1)M}{a_0} a_1 a_{-1}
-\frac{(2 a_1+b_1) N}{b_0} b_1 b_{-1} \right)
+\cO(\eta_0^2)\,,
\end{split}
\ee
\be
\begin{split}
e_{-1}^{(1;1)}=&4 \sqrt3 \epsilon \left[a_0 \left(a_{-1}  M
+b_{-1}( M-N)\right)+b_0 \left( a_{-1} (M-N)-b_{-1}N \right)\right]
\\
&-2\sqrt3 \epsilon \left( \frac{(a_{-1}+2b_{-1})M}{a_0} a_1 a_{-1}
-\frac{(2 a_{-1}+b_{-1}) N}{b_0} b_1 b_{-1} \right)
 +\cO(\eta_0^2)\,.
\end{split}
\ee

\end{document}